\documentclass[useAMS,usenatbib]{mn2e}
\usepackage{times}
\usepackage{natbib}
\usepackage{aas_macros}
\usepackage{amsmath}
\usepackage{amssymb}
\usepackage[pdftex]{graphicx}

\title[]{An Efficient Fokker-Planck Solver and its Application to Stochastic Particle Acceleration in Galaxy Clusters}
\author[J. Donnert, G. Brunetti]{
J. Donnert$^{1}$,\thanks{donnert@ira.inaf.it}
G. Brunetti$^{1}$\\
$1$INAF Istituto di Radioastronomia, via P. Gobetti 101, I-40129 Bologna, Italy\\
}

\begin{document}

\date{Accepted ???. Received ???; in original form ???}

\pagerange{\pageref{firstpage}--\pageref{lastpage}} \pubyear{2011}

\maketitle

\label{firstpage}

\begin{abstract}
Particle acceleration by turbulence plays a role in many astrophysical environments. The non-linear evolution of the underlying cosmic-ray spectrum is complex and can be described by a Fokker-Planck equation, which in general has to be solved numerically. We present here an implementation to compute the evolution of a cosmic-ray spectrum coupled to turbulence considering isotropic particle pitch-angle distributions and taking into account the relevant particle energy gains and losses. 
Our code can be used in run time and post-processing to very large  astrophysical fluid simulations. We also propose a novel method to compress cosmic-ray spectra by a factor of ten, to ease the memory demand in very large simulations.  We show a number of code tests, which firmly establish the correctness of the code. In this paper we focus on relativistic electrons, but our code and methods can be easily extended to the case of hadrons.
We apply our pipeline to the relevant problem of particle acceleration in galaxy clusters. We define a sub-grid model for compressible MHD-turbulence in the intra-cluster-medium and calculate the corresponding reacceleration timescale from first principles. We then use a magneto-hydrodynamic simulation of an isolated cluster merger to follow the evolution of relativistic electron spectra and radio emission generated from the system over several Gyrs.  

\end{abstract}


\section{Introduction}\label{sect.intro}
Most of the diffuse baryonic matter in the Universe is ionised and resides in very low-density environments. These underdense plasmas exhibit nearly infinite conductivity and negligible viscosity on collisional scales \citep[e.g. ][]{2004ppa..book.....K}. Given the strong driving forces present in many systems, this makes turbulence ubiquitous in space and astrophysical flows. Remarkably, turbulence itself affects the micro-physics of the medium \citep[e.g.][]{2010MNRAS.405..291S,2012SSRv..173..557L,2014mcp..book...87B}. Hence it is intimitely connected to the micro-physical and macro-physical properties of the background astrophysical plasma. Turbulence can also trigger several mechanisms of particle acceleration through the interaction of electromagnetic fluctuations in the plasma with thermal or relativistic particles. This nonlinear interplay between particles and turbulent plasma modes is a stochastic process that transfers energy from large-scale bulk motions into relativistic particles. \par

The interaction of turbulence and relativistic particles is an indispensible component of models of cosmic-ray (CR) propagation and acceleration \citep[e.g.][]{1949PhRv...75.1169F,1966SvA.....9..877G,1966ApJ...143..961J,1969pia..conf..271W}. Particle acceleration by MHD turbulence is a fairly robust process that is considered important for solar flares, $\gamma$-ray bursts and many other astrophysical environments \citep[e.g.][]{1992ApJ...398..350H,1996ApJ...461..445M,2000A&A...360..789S,2001ApJ...556..479D,2004ApJ...614..757Y,2010ApJ...720..503C,2012SSRv..173..535P,2011PhRvL.107i1101M}.\par

In galaxy clusters, the intra-cluster-medium (ICM) is a prototypical example for the aforementioned turbulent flow, where turbulence plays a role for particle acceleration and propagation. Here, the gravitational potential and approximate hydrostatic equilibrium imply high temperatures and low densities of the baryonic component \citep{1988xrec.book.....S}. Hierarchical structure formation predicts the late formation of clusters by constant in-fall and merging of halos, which drive shocks and turbulent motions on a wide range of scales in the ICM \citep{2004A&A...426..387S,2005MNRAS.357.1313C,2006MNRAS.366.1437S,2009A&A...504...33V,2011ApJ...726...17P,2012A&A...544A.103V,2014ApJ...782...21M}. This merger-induced turbulence is likely reaccelerating relativistic electrons, which power giant radio halos \citep[][]{2001MNRAS.320..365B,2001ApJ...557..560P,2014IJMPD..2330007B}. \par

The interaction between particles and turbulent fluctuations is usually modelled using quasi-linear theory (QLT), where the effect of linear waves on particles is studied by calculating first-order corrections to the particle orbit in the uniform/background magnetic field, and then ensemble-averaging over the statistical properties of the turbulent modes. These wave-particle interactions affect particle diffusion and transport through pitch-angle scattering. This has been  studied in many astrophysical systems, e.g. diffusion of CRs in our galaxy \citep[e.g.][ for a recent review]{2013A&ARv..21...70B}, and its role in the process of particle acceleration in shocks \citep[][ for a review]{1987PhR...154....1B}. The resulting particle energy gain is a slower process, with a typical acceleration time-scale $\tau_\mathrm{acc} \approx \tau_\mathrm{sca} (c/V_\mathrm{ph})^2$ where $\tau_\mathrm{sca}$ is the pitch-angle scattering time-scale and $V_\mathrm{ph}$ the phase-velocity of the waves. The acceleration by turbulent low-frequency MHD waves has been studied extensively, especially for phase velocities $V_\mathrm{ph} \ll c$. In this case, the magnetic field component associated with the waves is much larger than the electric field component, $\delta B \sim \delta E c/V_{ph}$, and the pitch-angle distribution of particles is quasi-isotropic. However, even in this simplified case the particle distribution function evolves according to Fokker-Planck equations that in general cannot be solved analytically and require complex numerical algorithm to obtain fast and stable solutions; this is the main focus of our paper\par

A seminal study on the analytical and numerical properties of solutions of isotropic Fokker-Planck equations has been presented by \citet{1995ApJ...446..699P,1996ApJS..103..255P} and has been recently extended by \citet{2011JCAP...12..010M}. In astrophysics, numerical solutions have been obtained in numerous studies in several fields \citep[e.g.][]{1986ApJ...308..929B,1996ApJ...461..445M,2001APh....15..223B,2004MNRAS.350.1174B,2005MNRAS.357.1313C,2006ApJ...647..539B,2011ApJ...739...66T,2011MNRAS.410..127B,2007MNRAS.378..245B,1995ApJ...452..912M,2006ApJ...636..798L,2004ApJ...610..550P}, whereas stochastic CR acceleration has been considered in the context of full astrophysical fluid simulations only recently \citep[see ][ for cluster simulations]{2013ApJ...762...78Z,2013MNRAS.429.3564D}.\par

We present here an  efficient, memory conserving formalism to follow and evaluate the acceleration and evolution of CR spectra in the context of high resolution astrophysical simulations. Our implementation is based on the \citet{1970CompPhys.ChangCooper} algorithm, which we extend to a numerical subgrid model for CR acceleration by turbulence. We begin with our particle acceleration model in section \ref{sect.reacc}, followed by details on the numerical algorithm and code tests in section \ref{sect.solver}. There we also present a novel algorithm for the compression of CR spectra. We then  apply our formalism to galaxy cluster simulations. In section \ref{sect.turb} we discuss turbulence in the collisionless ICM, in section \ref{sect.gadget} we show the application to MHD-SPH code {\small GADGET3}.  We draw our conclusions in section \ref{sect.concl}.

\section{Particle Acceleration Model} \label{sect.reacc}

The aim of the paper is to study the evolution of the spectral energy distribution of particles in a turbulent magnetised medium. We will focus on the case of CR electrons, with an isotropic distribution of pitch-angles. These particles provide important constraints to  transport and acceleration processes in astrophysical environments as they can be traced through their synchrotron and inverse Compton emission. Nonetheless the formalism can be extended to CR protons and to their secondary electrons. \par 

The interaction of CR electrons with turbulent MHD waves provides a stochastic scattering agent which energises particles. In addition, CR electrons are subject to a number of systematic energy losses, mainly due to the interaction with photons and magnetic fields, due to Coulomb and ionization collisions with thermal particles and due to compression and rarefaction of the background plasma in which they are embedded.  \par

The acceleration of particles through the stochastic interaction with magnetic turbulence is due to the interaction with the electric-field fluctuations associated with the waves. In general these fluctuations are much smaller than the magnetic field fluctuations that govern the process of particle pitch-angle scattering. As a consequence we shall assume that particles undergo an efficient isotropization and that the energy/momentum distribution function of CRs is \emph{isotropic}. Under these conditions, the equations governing the time evolution of an ensemble of CR electrons is \citep{1980panp.book.....M,1991bja..book..428E,2002cra..book.....S}:  

\begin{align}\label{eqn.reacc}
    \frac{\partial n(p,t)}{\partial t} &= \frac{\partial}{\partial p}\left[ n(p,t)\left( \sum_i \left|\frac{\mathrm{d}p}{\mathrm{d}t}\right|_{\mathrm{i}} -\frac{2}{p} D_\mathrm{pp}(p) \right) \right] \nonumber\\
	 &+ \frac{\partial}{\partial p}\left[ D_\mathrm{pp}(p) \frac{\partial n(p,t)}{\partial p} \right] + Q_\mathrm{e}(p,t) - \frac{n(p,t)}{T_\mathrm{e}(p,t)},
\end{align}
\noindent where $n(p,t) = 4 \pi p^2 f({\bf p},t)$ is the isotropic number density of CR electrons and we sum over all relevant energy losses covered in section \ref{sect.losses}. Equation \ref{eqn.reacc} takes the form a Fokker Planck equation, where turbulence provides systematic (first $D_{\mathrm{pp}}(p)$ term) and stochastic (second term) energy gain to the particle spectrum. Here, $Q_\mathrm{e}(p,t)$  describes particle injection and $T_\mathrm{e}$ catastrophic losses/escape from the system.

\subsection{Loss Mechanisms} \label{sect.losses}

\begin{figure}
	\centering
    \includegraphics[width=0.5\textwidth]{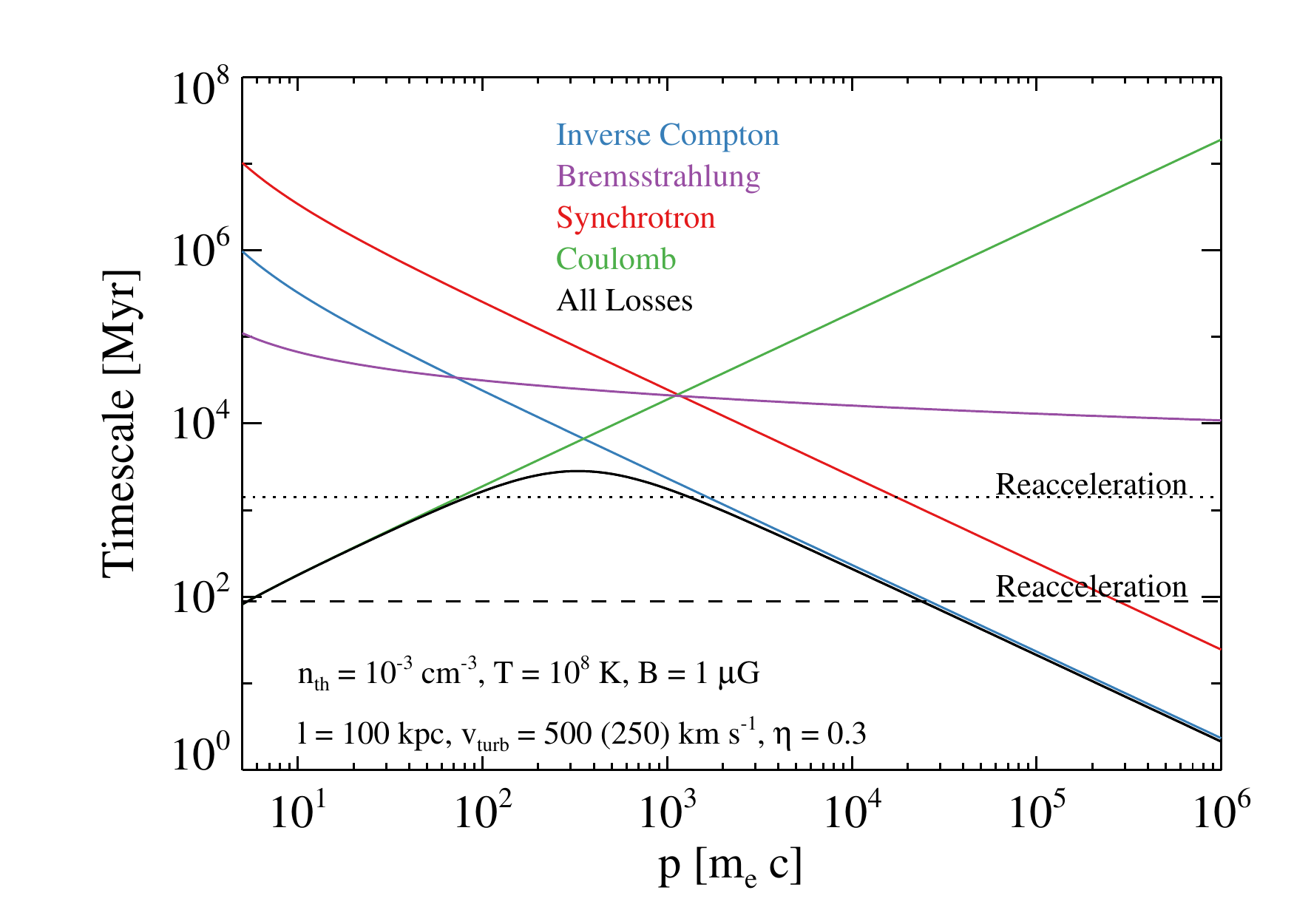}
	\caption{Cooling timescales in a galaxy cluster with $n_\mathrm{thermal} = 10^{-3} \mathrm{cm}^{-3}$, $T = 10^8 \,\mathrm{K}$, $B = 1 \,\mu\mathrm{G}$. Total cooling time from all losses (black line), from Coulomb losses eq. \ref{eq.ccoul} (green),  Bremsstrahlung losses eq. \ref{eq.cbrems} and from synchrotron and IC losses eq. \ref{eq.crad} (red and blue). We also overplot the reacceleration timescale $\tau_\mathrm{acc} = \frac{p^2}{4D_\mathrm{pp}}$ with $\eta = 0.3$ (eq. \ref{eq:Dpp_num}) for turbulence on a scale of 100 kpc with the velocities of 250 km/s (black dotted) and 500 km/s (black dashed). }\label{img.coolingtime}
\end{figure}

\subsubsection{Inverse Compton \& Synchrotron Losses}

For a magnetic field $B_{\mu G}$ in $10^{-6}\mathrm{G}$, synchrotron and inverse Compton losses can be modelled as \citep[][]{1986rpa..book.....R,1994hea2.book.....L}: 
\begin{align}\label{eq.crad}
     \left|\frac{\mathrm{d}p}{\mathrm{d}t}\right|_{\mathrm{IC}+\mathrm{Syn}}
	 &= \frac{4}{9} r_0^2  \beta^2\gamma^2  \left[ B_{\mu G}^2 + B_\mathrm{ph,
	 \mu G}^2 \left( 1+z\right)^4 \right], 
\end{align}
with the classical electron radius $r_0 = q_\mathrm{e}^2/m_\mathrm{e}/c^2$, and $B_\mathrm{ph, \mu G}$ the inverse Compton equivalent magnetic field due to the background radiation field  in micro Gauss.

\subsubsection{Coulomb Scattering}

Coulomb collisions of CR electrons with the ambient thermal protons and electrons  provide another source of energy losses. This can be estimated \citep{2002cra..book.....S} as:
\begin{align}\label{eq.ccoul}
	\left|\frac{\mathrm{d}p}{\mathrm{d}t}\right|_{\mathrm{CC}} &= \frac{4\pi r_0^2n_\mathrm{th}m_\mathrm{e}c^2}{\beta_\mathrm{e}} \ln\Lambda \\
    \ln\Lambda &= 37.8 + \log\left( \frac{T}{10^8} \left( \frac{n_\mathrm{th}}{10^3}\right)^{-1/2} \right),
\end{align}
where $n_\mathrm{th}$ is the thermal number density, $T$ the plasma temperature in Kelvin and $\ln\Lambda$ is the Coulomb logarithm at $T>4\times 10^5 \,\mathrm{K}$ \citep{1988xrec.book.....S}.

\subsubsection{Bremsstrahlung}

On scales of the Debye length CRe interact with the ambient mean electromagnetic field and emit free-free radiation. In a completely ionised medium with a Helium fraction of $X_\mathrm{He}$ and a hydrogen fraction $X_\mathrm{H}$ this is \citep{1970RvMP...42..237B}:
\begin{align}\label{eq.cbrems}
	\left|\frac{\mathrm{d}p}{\mathrm{d}t}\right|_{\mathrm{ff}} 	&= 8 \alpha
	r_0^2 m_\mathrm{e}c^2 n_\mathrm{th} \gamma \left[ X_\mathrm{H} +
	3X_\mathrm{He} \right] \left[ \ln(2\gamma) - \frac{1}{3} \right],
\end{align}
where $\alpha$ is the fine structure constant.

\subsubsection{Adiabatic Expansion}

The invariance of phase-space density of the CR population implies that a change in physical density of the medium changes the momentum spectrum of the CR particles. This is \citep{1960SvA.....4..243S,1962AZh....39..393K}:
\begin{align}
	\left|\frac{\mathrm{d}p}{\mathrm{d}t}\right|_{\mathrm{AdEx}} &= -\frac{1}{3} \left(\nabla \cdot {\bf v} \right) p, \label{eq.adexloss}
\end{align}
where ${\bf v}$ is the expansion velocity.

\subsection{Reacceleration Coefficient}

The reacceleration of CRe by turbulence can be described appropriately by
quasi-linear theory (QLT), under the condition of small amplitude of magnetic field fluctuations, $\delta B << B_0$ where $B_0$ is the background field. \par

Under negligible damping, the polarisation and dispersion relation of the MHD waves identify as the standard Alfven, slow and fast modes. Gyro-resonance is the strongest resonance condition between CRs and Alven waves, $\omega = k_\parallel \nu_\parallel + n \Omega / \gamma$, with $n=\pm 1$.  Fast
modes interact also via the \emph{Transit Time Damping} (TTD) mechanism that satisfies the condition $\omega = k_\parallel \nu_\parallel$, where $\nu_\parallel$ is the particle velocity along the field and $ k_\parallel$ the wave-number along the field, $\omega$ the wave frequency, and $\Omega = \frac{eB}{m_e c}$ the electron gyro-frequency. \par

The reacceleration coefficient is formally given by \citep[e.g. ][ eq. 1c]{1998ApJ...492..352S}:
\begin{align}
	D_\mathrm{pp} &= \lim\limits_{t\rightarrow\infty} \frac{1}{2t} \left<
	\delta {\bf p}(t) \delta{\bf p}^*(t+\tau)\right> \\
	 	&= \mathcal{R} \int \limits^{\infty}_{0} \,\mathrm{d}\tau \left<
	\dot{{\bf p}}(t)\dot{{\bf p}}^*(t+\tau)\right>
\end{align}
here $\delta {\bf p} \delta{\bf p}^*$ represent the change of momentum due to the interaction with electromagnetic fluctuations. Relevant coefficients can be found in \citet{2002cra..book.....S}. \par

A less general, but useful way to derive the reacceleration coefficient, is the argument of detailed balancing \citep[e.g.][]{1979ApJ...230..373E,1981A&A....97..259A}. Here one recognises, that the total energy added to all CRe with momentum in the range $p_{min}$ and $p_{max}$, is equal to the total energy loss of the turbulent wave spectrum $W({\bf k})$ in a corresponding range of wave-numbers due to CRe damping $\Gamma^\mathrm{e}(k, \theta)$. 
\begin{align}\label{eq.detailedb}
	\int\limits^{p_\mathrm{max}}_{p_\mathrm{min}}\mathrm{d}^3p \, E_\mathrm{e}
	\left( \frac{\partial f(p)}{\partial t} \right) =
	\int\limits^{k_\mathrm{max}}_{k_\mathrm{min}}  \mathrm{d}{\bf k}\, \Gamma^\mathrm{e}(k) W({\bf k}),
\end{align}

Assuming pitch-angle isotropy, the time evolution of $f(p)$ due to stochastic interactions with turbulence is: 
\begin{align}
	\frac{\partial f}{\partial t} &= \frac{1}{p^2} \frac{\partial}{\partial p} \left( p^2 D_\mathrm{pp} \frac{\partial f}{\partial p} \right),
\end{align}
which implies:
\begin{align}
	\int\limits^{p_\mathrm{max}}_{p_\mathrm{min}}\mathrm{d}^3p \,
	\frac{E_\mathrm{e}}{p^2} \frac{\partial}{\partial p} \left( p^2
	D_\mathrm{pp} \frac{\partial f}{\partial p} \right) =
	\int\limits^{k_\mathrm{max}}_{k_\mathrm{min}}  \mathrm{d}{\bf k}\, \Gamma^\mathrm{e}(k) W({\bf k}). \label{eq.isobalance}
\end{align}
Equation \ref{eq.isobalance} represents a valuable way to derive $D_\mathrm{pp}$ once the damping rate $\Gamma$ and the wave spectrum are known.  In several relevant cases, such as the TTD that will be used in Section 4, the damping rate depends on the (derivative of the) CR momentum distribution and takes the form:
\begin{align}
	\Gamma^\mathrm{e}({\bf k}) &= - \bar{\Gamma}({\bf k})
	\int\limits^{p_\mathrm{max}}_{p_\mathrm{min}} A(p)  \frac{\partial f(p)}{\partial p}  \, \mathrm{d}p, \label{eq.msdamp}
\end{align}
where $A(p)$ is a function of CR momentum and $\bar\Gamma(k)$ of the wavenumber. \par

By combining eq. \ref{eq.isobalance} with eq. \ref{eq.msdamp} and partial
integration of the former one obtains the expression for the reacceleration coefficient:
\begin{align} \label{eq.dppure}
	D_\mathrm{pp} &= \frac{m_\mathrm{e}}{8\pi} \frac{A(p)}{p^3}
	\sqrt{1+\left(\frac{p}{m_\mathrm{e}c}\right)^2}
	\int\limits^{k_\mathrm{max}}_{k_\mathrm{min}}  \bar{\Gamma}({\bf k}) W({\bf k}) \, \mathrm{d}^3 k
\end{align}

\section{Numerical Fokker-Planck Solver} \label{sect.solver}

\subsection{Algorithm} \label{sect.FKPalg}

We use the finite difference scheme proposed by \citet{1970CompPhys.ChangCooper}. Using no-flux boundary conditions, it ensures positivity, particle conservation and correctness of the equilibrium solution by construction. Furthermore it is well suited for logarithmic grids and unconditionally stable, which allows us to use a comparatively small number of gridpoints and speed-up the calculation. 
Equation \ref{eqn.reacc} can be written as: 
\begin{align} \label{eqn.numfkp}
    \frac{\partial n(p)}{\partial t} &= \frac{\partial}{\partial p} \left[ H(p) n(p)  + D_{\mathrm{pp}} \frac{\partial n(p)}{\partial p} \right] \nonumber\\ 
    & -\frac{n(p)}{T_{\mathrm{e}}(p,t)} + Q_{\mathrm{e}}(p,t)
\end{align}
We introduced the generalised cooling function $H(p)$ for simplicity: 
\begin{align}
    H(p) &= -\frac{2}{p}D_{\mathrm{pp}} + \sum_i \left|\frac{dp}{dt} \right|_{i}.
\end{align}
Here we sum over all relevant loss functions for the particle population (see section \ref{sect.losses}).\par

We use a logarithmic momentum grid with N points, so for $i \in \{0...N-1\}$:
\begin{align}
    p_{\mathrm{i}} &= p_{\mathrm{min}}  10^{i p_{\mathrm{step}}}, \\
	p_{\mathrm{step}} &= \log_{10} \left(\frac{p_{\mathrm{max}}}{p_{\mathrm{min}}}\right) \left( N-1 \right)^{-1}  \\
    \Delta p_{i} &= (p_{i+1} - p_{i-1})/2\\
    \Delta p_{i+1/2} &= p_{i+1}-p_{i},
\end{align}
so that for a quantity on the grid X(p): 
\begin{align}
    X_{i} &= X(p_{i}),& X_{i+\frac{1}{2}} &= \frac{1}{2} (X_{i} + X_{i+1})
\end{align}
and the time dependence:
\begin{align}
    n_{i}^{j} &= n(p_{i}, t_{j}) .
\end{align}
Following \citet{1996ApJS..103..255P},  equation \ref{eqn.numfkp} can be expressed in terms of fluxes as: 
\begin{align} \label{eqn_fkp_descr}
    \frac{n_{i}^{j+1} - n_{i}^{j}}{\Delta t} &= \frac{F_{i+1/2}^{j+1}-F_{i-1/2}^{j+1}}{\Delta p_{i}} - \frac{n_{i}^{j+1}}{T_{i}} + Q_{i}
\end{align}
The Chang \& Cooper method exploits the degree of freedom introduced when discretising the cooling function in the flux $F_{i}^{j}$ so the method is adaptively 'upwind', towards the steady-state solution: 
\begin{align}
    F^{j+1}_{i+1/2} &= (1 - \delta_{i+1/2}) H_{i+1/2}n_{i+1}^{j+1} + \delta_{i+1/2}H_{i+1/2}n_{i}^{j+1} \nonumber\\&+ D_{\mathrm{pp}, i+1/2}\frac{n_{i+1}^{j+1} - n_{i}^{j+1}}{\Delta p_{i+1/2}} \nonumber\\
    &= H_{i+1/2} \left[W_{i+1/2}^{+} n_{i+1}^{j+1} - W_{i+1/2}^{-} n_{i}^{j+1}  \right] \label{eqn_fkp_flux2}
\end{align}
with
\begin{align}
    \delta_{i} &= \frac{1}{w_{i}} - \left[ e^{w_{i}} -1 \right]^{-1}, & w_{i}  &= \frac{H_{i}}{D_{\mathrm{pp},i}} \Delta p_{i} \\
    W^{+}_{i}  &= \frac{1}{1- e^{-w_{i}}},& W^{-}_{i} &= \frac{1}{e^{w_{i}}-1},
\end{align}
which can be calculated very efficiently. To avoid floating point over/underflow in the coefficients, we limit $w_i \in [10^{-8}, 700]$ in the code without loss of accuracy. \par
Combining equations \ref{eqn_fkp_descr} and \ref{eqn_fkp_flux2} leads to a tridiagonal system of equations :
\begin{align}
    -A_{i}n^{j+1}_{i+1} + B_{i}n^{j+1}_{i} + C_{i}n^{j+1}_{i-1} &= r_{i}\\
    A_{N-1} = C_{0} &= 0 
\end{align}
with
\begin{align}
    A_{i} &= \frac{\Delta t}{\Delta p_{i}} H_{i+1/2} W^{+}_{i+1/2} \\
    B_{i} &= 1 + \frac{\Delta t}{\Delta p_{i}} \left[ H_{i-1/2} W^{+}_{i-1} + H_{i+1/2} W^{-}_{i+1} \right] + \frac{\Delta t}{T_{i}}\\
    C_{i} &= \frac{\Delta t}{\Delta p_{i}} H_{i-1/2} W^{-}_{i-1/2} \\
    r_{i} &= \Delta t Q_{i} + n_{i}^{j}.
\end{align}

The tridiagonal system is then solved using a tridiagonal matrix solver \citep[e.g.][]{1992nrfa.book.....P}. \par

\subsection{Time Stepping}

An estimate of the evolution time scale can be obtained from the cooling timescale:  
\begin{align}
	\tau(p_k) &= \frac{p_k}{H(p_k)},
\end{align}
at all momenta $p_k$. We take:
\begin{align}
	\Delta \tau &= \frac{1}{2} \min\left(t(p_k)\right),
\end{align}
where we only consider $k$ at which the spectrum is nonzero. We employ individual time steps similar to the block time step scheme used in N-body codes  \citep{1991PASJ...43..859M,2011EPJP..126...55D}.

\subsection{Boundary Conditions}

Proper treatment of boundary conditions is essential for the analytic as well as the numeric solution of Fokker-Planck type equations. Under no-flux boundary conditions, the Chang \& Cooper method can be shown to conserve the particle number \citep[e.g.][]{1996ApJS..103..255P}, which makes this choice useful for code tests. The no-flux condition is equivalent to vanishing Fokker-Planck coefficients at the two momentum bins next to the boundary. \par
However for our application we need to allow flux through the low momentum boundary, to avoid an unphysical pile-up of particles at the domain edge. This can be done , by truncating the spectrum at a momentum $p_{\mathrm{cut}}$ near the boundary and extrapolating $n(p)$ up to the boundary \citep{1986ApJ...308..929B}. I.e. every timestep we set $n(p<p_{\mathrm{cut}}) = 0$ and then extrapolate $n(p)$ for a few momentum bins as a power-law, based on the spectrum at $p>p_\mathrm{cut}$. The 'buffer' region usually spans around 5-10 grid points, depending on the problem.

\subsection{Code Tests} \label{sect.code_tests}

\subsubsection{Hard Sphere Equations}
\begin{figure*}
    \includegraphics[width=0.49\textwidth]{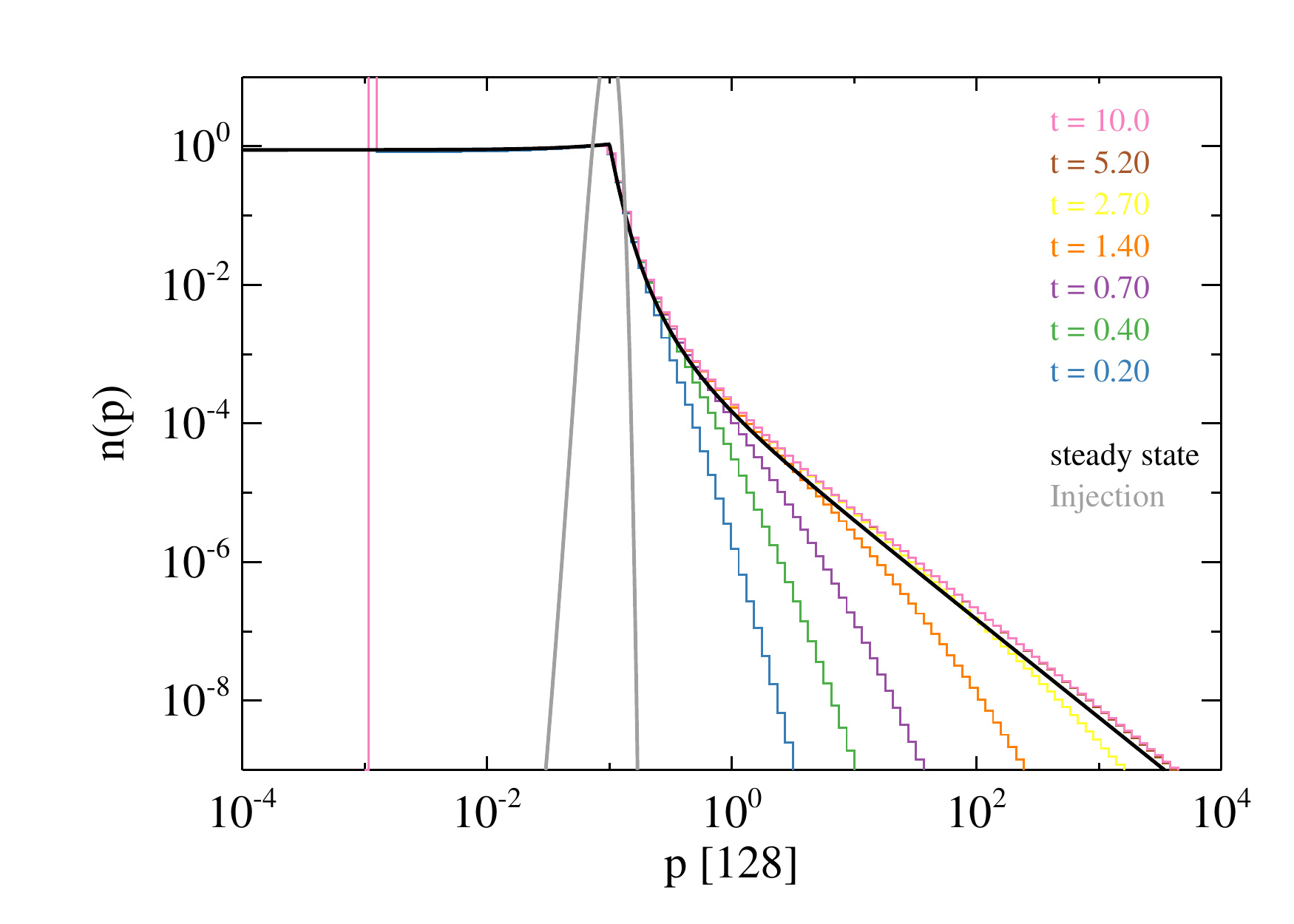}
    \includegraphics[width=0.49\textwidth]{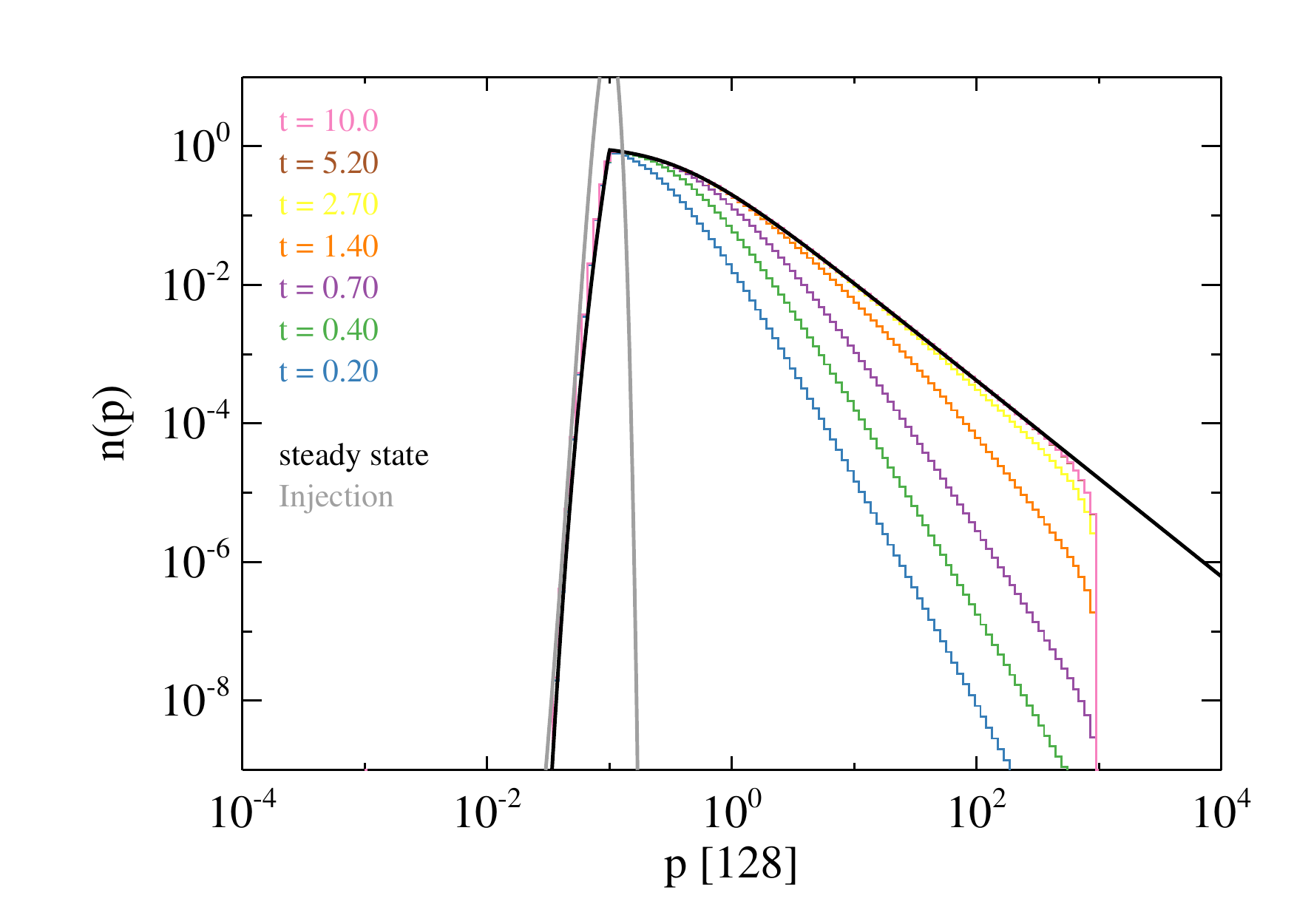} \\
    \includegraphics[width=0.49\textwidth]{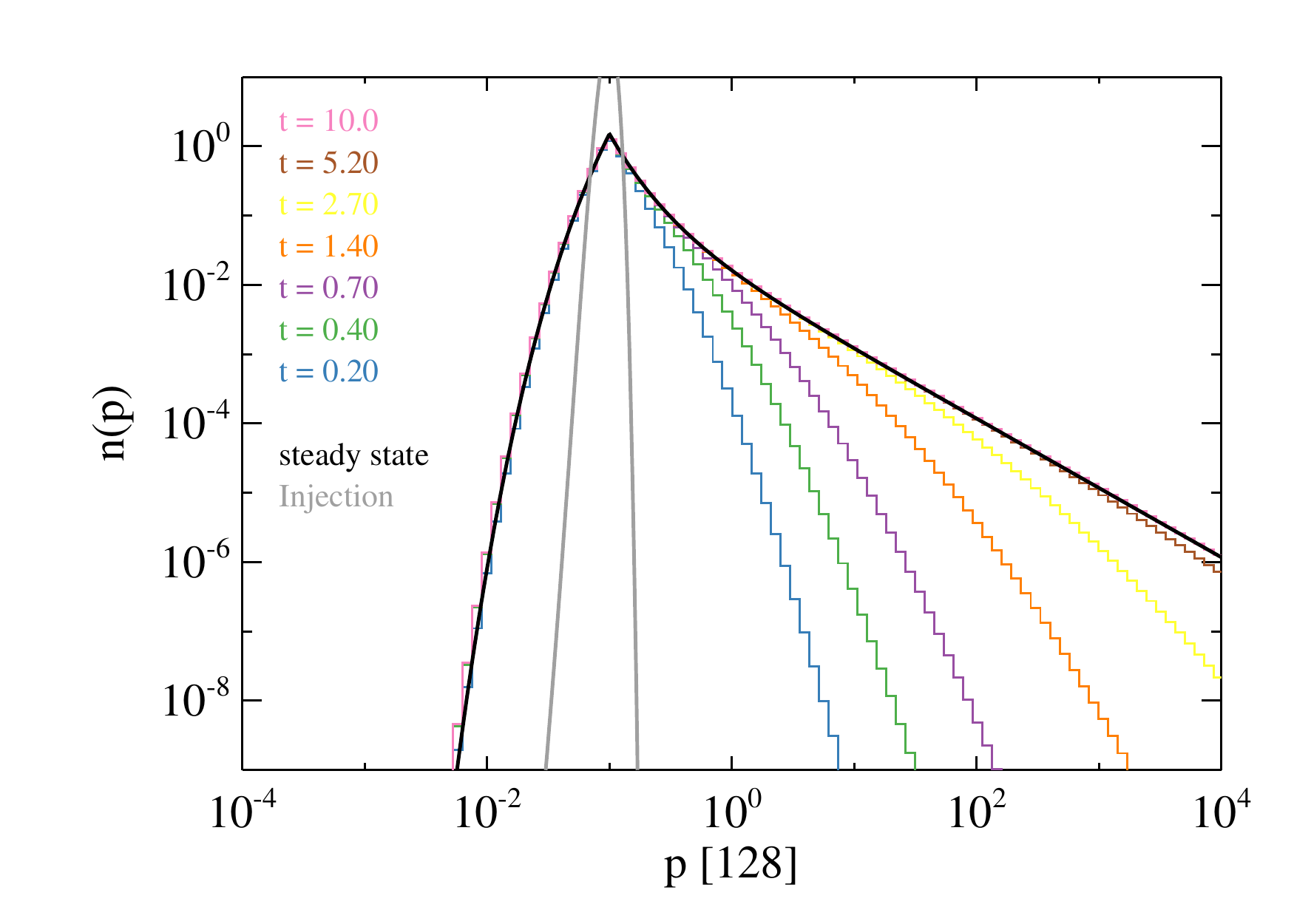}
    \includegraphics[width=0.49\textwidth]{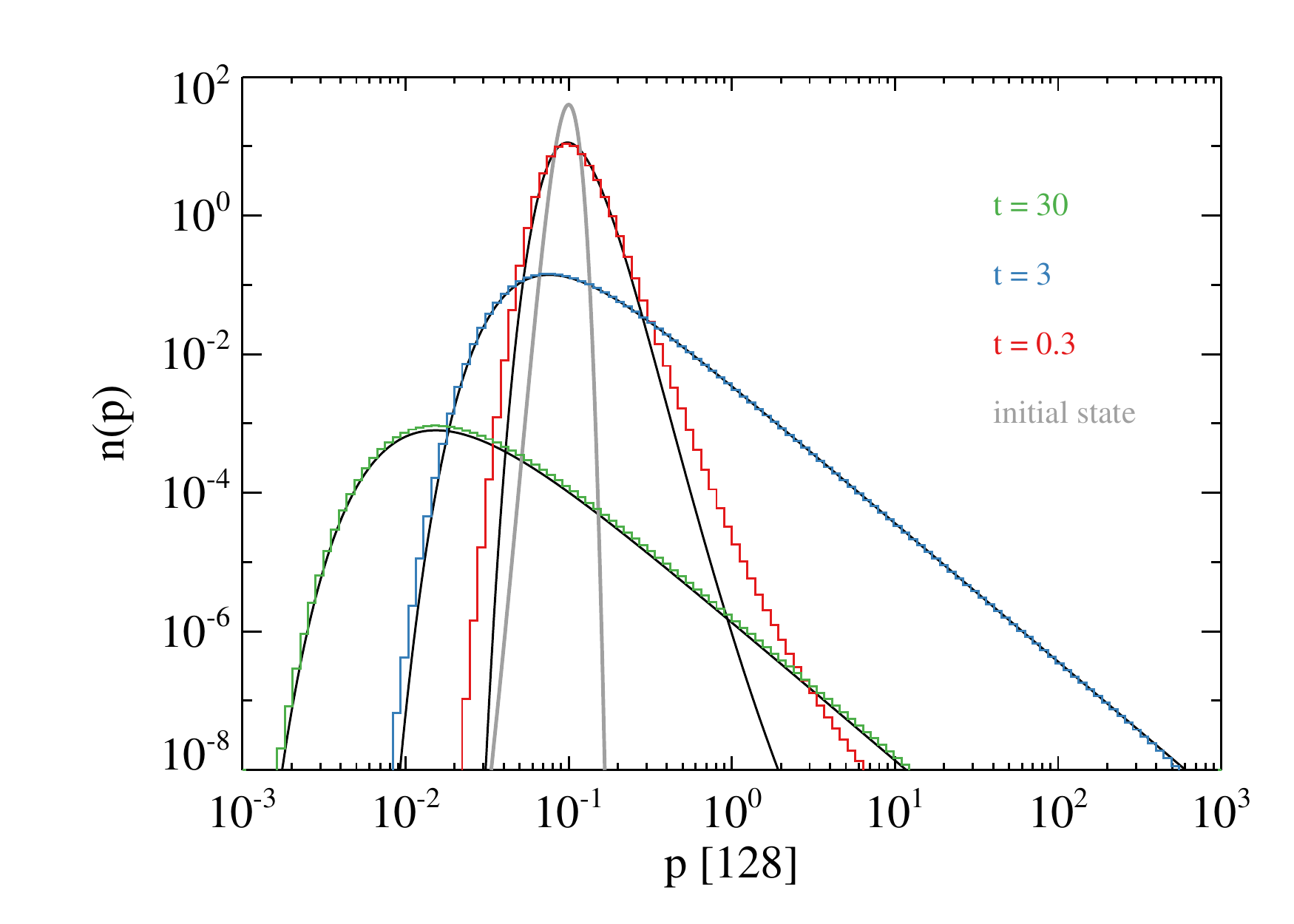}
	\caption{Analytical (black line) and numerical solutions (colors) to the hard-sphere Fokker-Planck equations \ref{eq.hs} ($a=1$ and $a=-1$), \ref{eq.hs2} and \ref{eq.hs3}, top left to bottom right. The injection spectrum is shown in gray. Boundary effects are can be seen  at the top left for low momenta  and at the top right for high momenta. In the bottom right panel, the curve at $t=30$ has been multiplied by $10^{11}$ for clarity.}\label{img.hardsphere}
\end{figure*}
To test our implementation it is useful to compare its numerical solution to analytic solutions of the Fokker-Planck equation. Unfortunatelly general time dependent solutions to equation \ref{eqn.reacc} are not known. However there have been extensive studies on a specific type of Fokker-Planck equation \citep{1970PhRvD...2.2787J,1984A&A...136..227S,1995ApJ...446..699P,2011JCAP...12..010M}. In the "hard-sphere approximation" the interaction of a thermal plasma with magnetic turbulence is described similar to elastic collisions between hard spheres considering Coulomb and Fermi interactions \citep{1958PhRv..111.1206P}. We follow the analysis of \citet{1995ApJ...446..699P,1996ApJS..103..255P} to test our code against a number of different Fokker-Planck equations. \par
The first hard sphere equation is given by:
\begin{align}\label{eq.hs}
    \frac{\partial n(x)}{\partial t} &= \frac{\partial}{\partial x} \left[ x^2 \frac{\partial n(x)}{\partial x} - n(x) - a x n(x) \right] \nonumber\\
    &- n(x) + \delta(x-x_0)\Theta(t),
\end{align}
with Heavyside step function $\Theta(t)$ and injection at momentum $x_0$. The steady state solution can be obtained from \citet{1988SoPh..115..313S, 1995ApJ...446..699P}. Following the latter we set $y = 1/x$ and for $a>0$:
\begin{align}
	n_\mathrm{ss}(x) &= \frac{\Gamma(\mu-\frac{a+1}{2})}{\Gamma(1 + 2\mu)} e^{-y_0} y_0^{\mu+(a+1)/2} y^{\mu+(1-a)/2} \Psi(x),
\end{align}
where
\begin{align}
	\Psi(x<x_0) &=  U\left(\mu+\frac{1-a}{2}, 1+2\mu, y \right) \nonumber\\ 
    &\times 1F1\left( \mu+\frac{1-a}{2}, 1+2\mu, y_0 \right)\label{eq.hs_1}  \\
	\Psi(x>x_0) &=  U\left(\mu+\frac{1-a}{2}, 1+2\mu, y_0 \right) \nonumber\\ 
    &\times 1F1\left( \mu+\frac{1-a}{2}, 1+2\mu, y \right)\label{eq.hs_2}
\end{align}
and
\begin{align}
    \mu &= \left[ 1+\frac{(3+a)^2}{4} \right]^{\frac{1}{2}}.
\end{align}
$U(\alpha,\beta,x)$ and $1F1(\alpha,\beta,z)$ are the confluent hypergeometric functions \citep{1970hmfw.book.....A}. \par
For $a<0$ the solution of \ref{eq.hs} is given by multiplying equations \ref{eq.hs_1} and \ref{eq.hs_2} by $\exp(y/y_0)$. \par
In figure \ref{img.hardsphere} top row, we plot the analytical solutions for $a=1$ (left) and $a=-1$ (right) in black, as well as the numerical solution at different times in colors. Everywhere we use a constant timestep of $\Delta t = 10^{-4}$. We add the injection function in gray. For late times our code agrees well with the steady stade solution. Boundary effects are limited to a few grid cells. \par
The next test case involves a hard sphere equation with strongly varying escape function: 
\begin{align}\label{eq.hs2}
    \frac{\partial n(x)}{\partial t} &= \frac{\partial}{\partial x} \left[ x^2 \frac{\partial n(x)}{\partial x}  - a x n(x) \right] \nonumber\\
    &- \frac{n(x)}{x} + \delta(x-x_0)\Theta(t).
\end{align}
The stationary solution is given in \citet{1995ApJ...446..699P} to be:
\begin{align}
	n(x) &= \frac{1}{|\alpha|} x^{(-1+a)/2}x_0^{(-1-a)/2} \Phi(y)
\end{align}
where
\begin{align}
	\Phi(y<y_0) &= I_\mu(y) K_\mu(y_0)\\
    \Phi(y>y_0) &= I_\mu(y_0) K_\mu(y)
\end{align}
where $I_\mu(y)$ and $K_\mu(y_0)$ are the Bessel functions and $\alpha = -1/2$, $y=x^\alpha$ and here $\mu=(1+\alpha)/(2\alpha)$. \par
For $a=1$ we plot the steady state solution in figure \ref{img.hardsphere} bottom left, in black, with the injection in gray and the time evolution of the numerical solution in color. Good agreement is found for the steady state, as expected from the Chang and Cooper method. \par
Finally we test the accuracy of the time evolution in the code. We consider the hard sphere equation:
\begin{align}\label{eq.hs3}
     \frac{\partial n(x)}{\partial t} &= \frac{\partial}{\partial x} \left[ x^3 \frac{\partial n(x)}{\partial x}  - a x^2 n(x) \right] \nonumber\\
    &- n(x) + \delta(x-x_0)\delta(t).
\end{align}
The time-dependent solution is given by \citep{1995ApJ...446..699P}:
\begin{align}
    n(p) &=  \frac{1}{2 \left|\alpha\right| t} x^{(-2+a)/2}  x_0^{(-2-a)/2} I_\nu\left( \frac{x^\alpha x_0^\alpha}{2\alpha^2 t} \right) \times \nonumber \\
    &\times\exp\left( -\frac{x^{2\alpha}+x_0^{2\alpha}}{4\alpha^2 t} \right)
\end{align}
with $\nu = \left| \frac{q-1+a}{2\alpha}\right|$, $a=1$, $q=2$, $s=0$, $x_0=0.1$, $\alpha=\frac{1}{2}(2-q-s)$. \par
The time dependent solution is shown for $t=0.3,3,30$ in figure \ref{img.hardsphere} alongside the initial spectrum in gray. We overplot the numerical solution in red, blue and green respectively. The curves at $t=30$ are multiplied by $10^{11}$. Good agreement is found at late times when the spectrum varies slowly. However at early time, when the spectrum is decaying quickly the numerical solution overestimates the correct solution in the outer parts of the spectrum. This is due to the low order of the Chang and Cooper method, it is quite diffusive. This has been noticed in \citet{1996ApJS..103..255P} as well. 

\subsubsection{Adiabatic Expansion}

\begin{figure*}
    \includegraphics[width=0.49\textwidth]{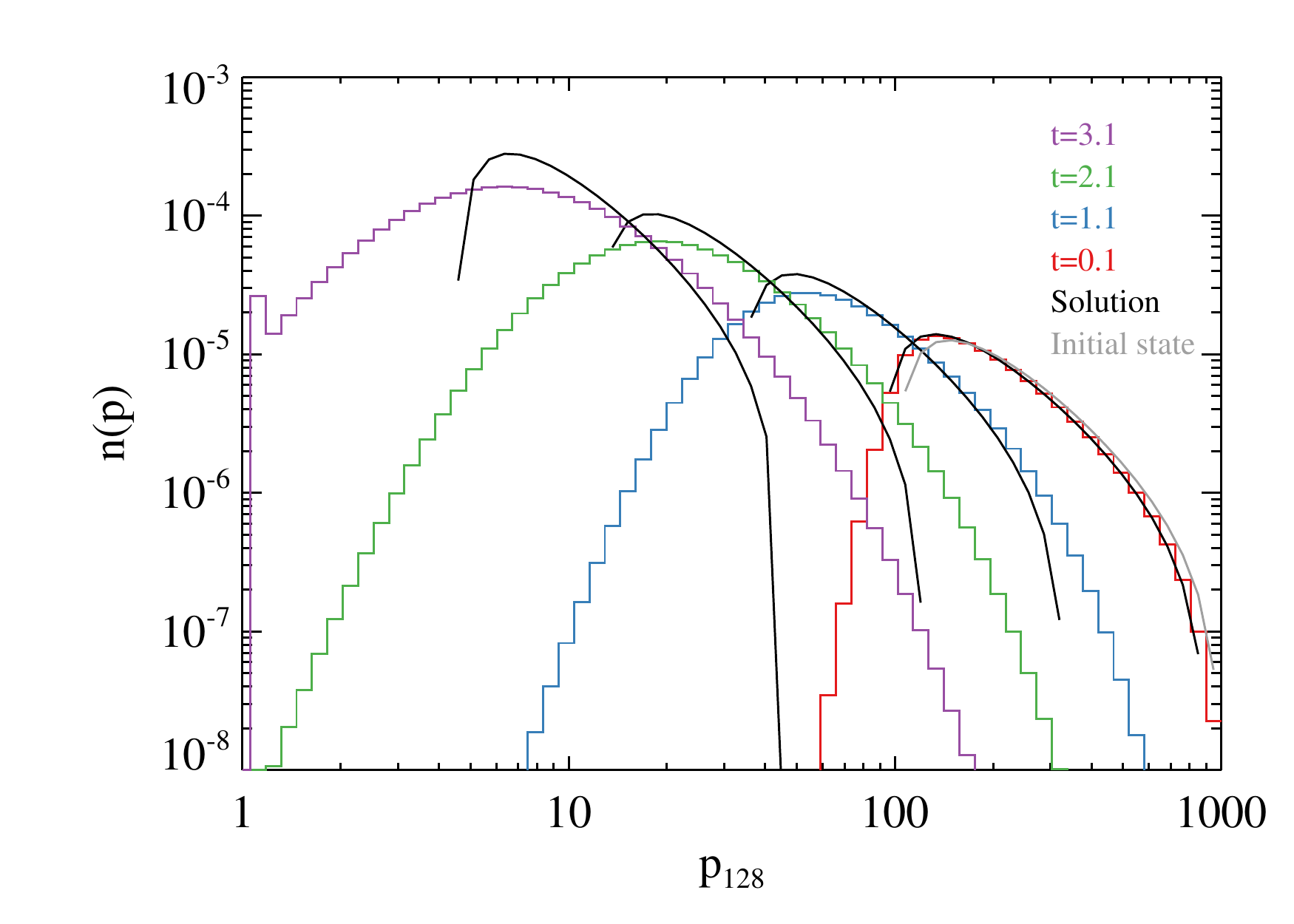}
    \includegraphics[width=0.49\textwidth]{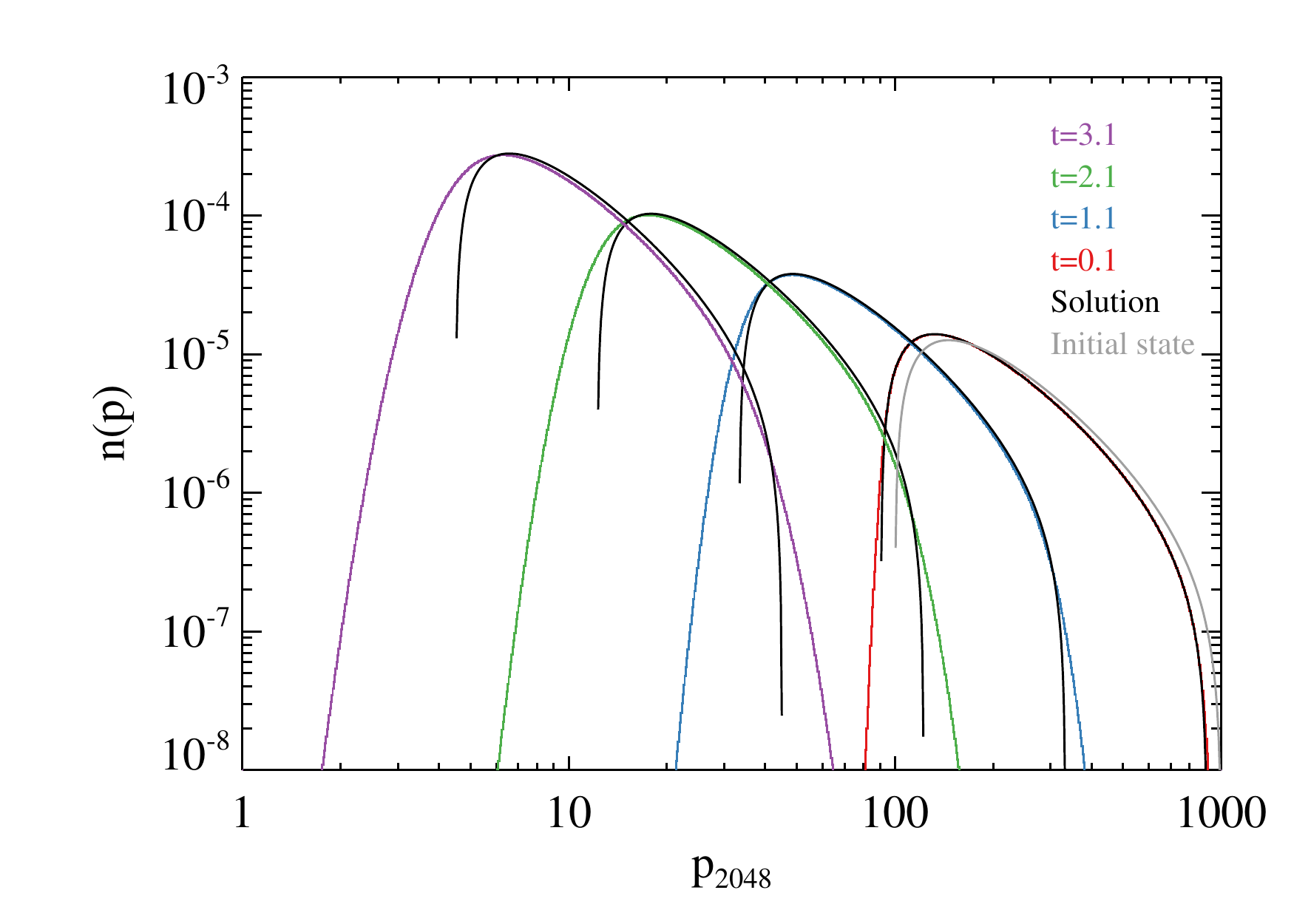}
	\caption{Time evolution of the \emph{comoving} solution (black, equation \ref{eq.diffsol}) to the diffusion equation \ref{eq.diff} and two numerical solutions with 128 (left) and 2049 (right) gridpoints. The initial spectrum is show in light grey. }
	\label{img.expansion}
\end{figure*}

The problem of diffusivity can be shown strikingly by considering the standard diffusion equation for constant adiabatic expansion, which is equivalent to $D_\mathrm{pp} = 0$ in equation \ref{eqn.numfkp} and considering only expansion losses through change in density (eq. \ref{eq.adexloss}):
\begin{align}
    \frac{\partial n(p)}{\partial t} &= \frac{\partial}{\partial p} \left[ \left( -\frac{1}{3} \frac{\dot{\rho}}{\rho}p\right)\, n(p)  \right] + Q(p), \label{eq.diff}
\end{align}
where we assume:
\begin{align}
	Q(p) &= N_0p^{-\gamma} \left(1-\frac{p}{p_\mathrm{HC}}\right) \Theta(p+p_\mathrm{LC}) \nonumber\\
	 &\left(1-\frac{p_\mathrm{LC}}{p} \right) \Theta(p_\mathrm{HC}-p), \label{eq.cassInjFunc}
\end{align}
where $\gamma = 2$ and $p_\mathrm{LC}$ and $p_\mathrm{HC}$ are the injection cut-off momenta.  The solution for the spectrum \emph{per comoving volume} can be found by the method of characteristics given the injection:
\begin{align}
	\hat{n}(p) &= N_0 p^{-\gamma} e^{\alpha (t-t_0)}\times \nonumber\\
   &\times\left[1 - \frac{p}{p_\mathrm{HC}} e^{-\alpha(t-t_0)}\right] \left[1 - \frac{p_\mathrm{LC}}{p} e^{\alpha(t-t_0)}\right], \label{eq.diffsol}
\end{align}
with $\alpha = -\dot{\rho}/3\rho$ and the spectrum per unit volume is $n(p,t)=\hat{n}(p,t) V(t_0)/V(t)$. \par
We plot this analytic solution in figure \ref{img.expansion} (black, dashed) alongside the numerical solution (colors) for constant $\alpha$ at different times. In the left panel we show a calculation using 128 grid-points on the right 4096 gridpoints. The diffusive behaviour for large grid-spacings is clearly visible in the low resolution run. 

\subsubsection{Convergence}
\begin{figure}
    \includegraphics[width=0.49\textwidth]{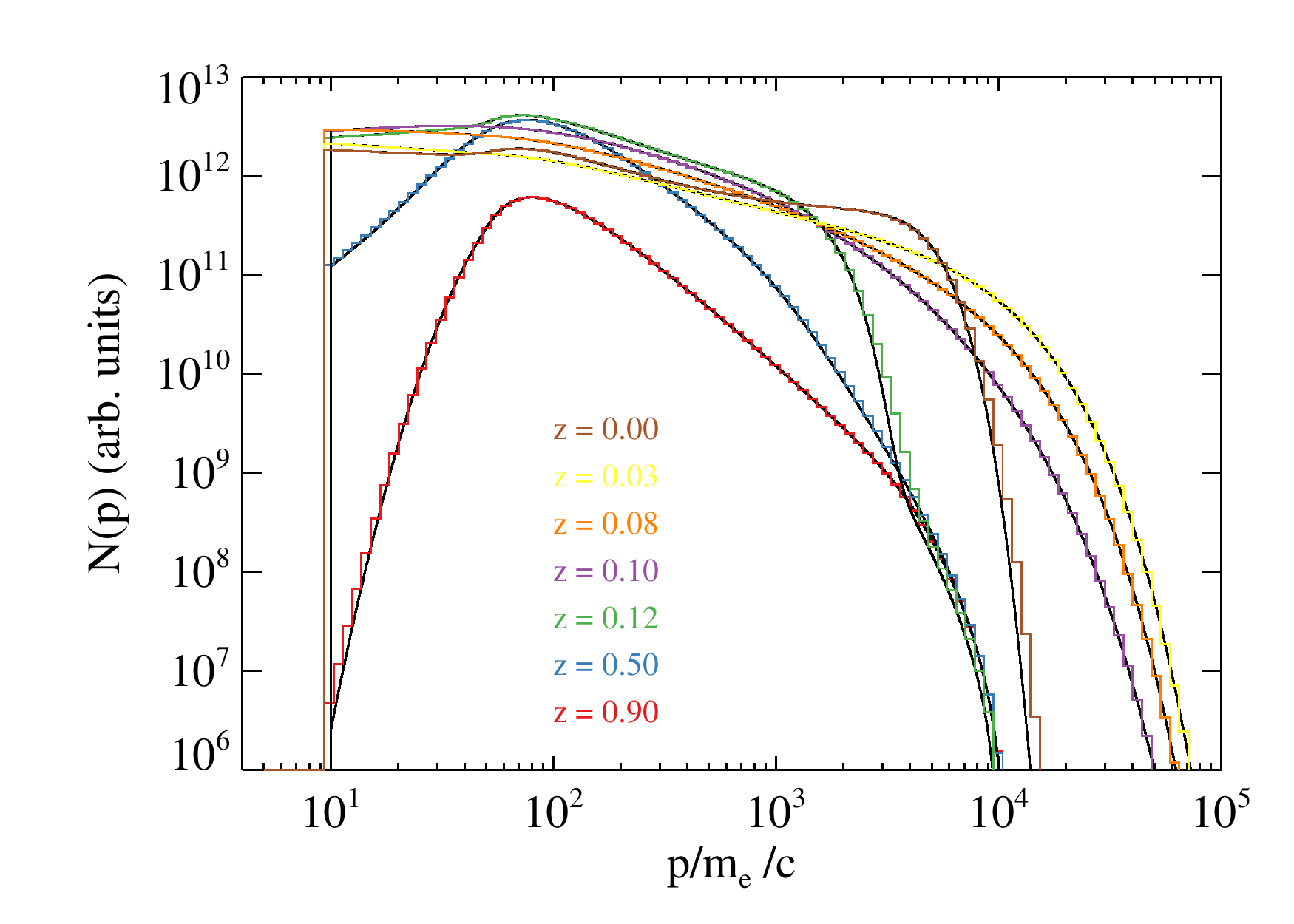}
    \caption{CR electron spectra using the time-dependent acceleration coefficient calculated by \citet{2005MNRAS.357.1313C} for one of their simulation runs. We show 2048 (black) and 128 grid points (colors) at different times.}\label{img.convergence}
\end{figure}

We test our code in a more complex / realistic situation where turbulent acceleration, synchrotron, inverse Compton and Coulomb losses are taken into account. We study convergence using the acceleration coefficient calculated by \citet{2005MNRAS.357.1313C} in simulated clusters.   To this end we assume $B = 1 \,\mu\mathrm{G}$, $n_\mathrm{th} = 10^{-4}$ and the loss functions given in section \ref{sect.losses}. \citet{2005MNRAS.357.1313C} used the extended Press-Schechter formalism to estimate the turbulent energy released in mergers over redshift. From this they obtained the redshift evolution of a cluster-wide reacceleration coefficient. The $D_\mathrm{pp}$ for a particular cluster in their simulations is given in table \ref{tab.cassanoDpp} in the appendix. The simulation runs from $z=1$ to $z=0$ with the constant injection function equation \ref{eq.cassInjFunc}, with  $p_\mathrm{HC} = 5000 \, m_\mathrm{e}c$ and $p_\mathrm{LC} = 100\, m_\mathrm{e}c$. We used open boundary conditions. \par
We show resulting spectral evolution for 2048 grid points (black) and 128 points (colors) in figure \ref{img.convergence}. Good agreement is found between the two resolutions. Possible deviations occur at steep gradients, because the lower resolution is more diffusive.

\subsection{Data Compression}
\begin{figure*}
    \includegraphics[width=0.49\textwidth]{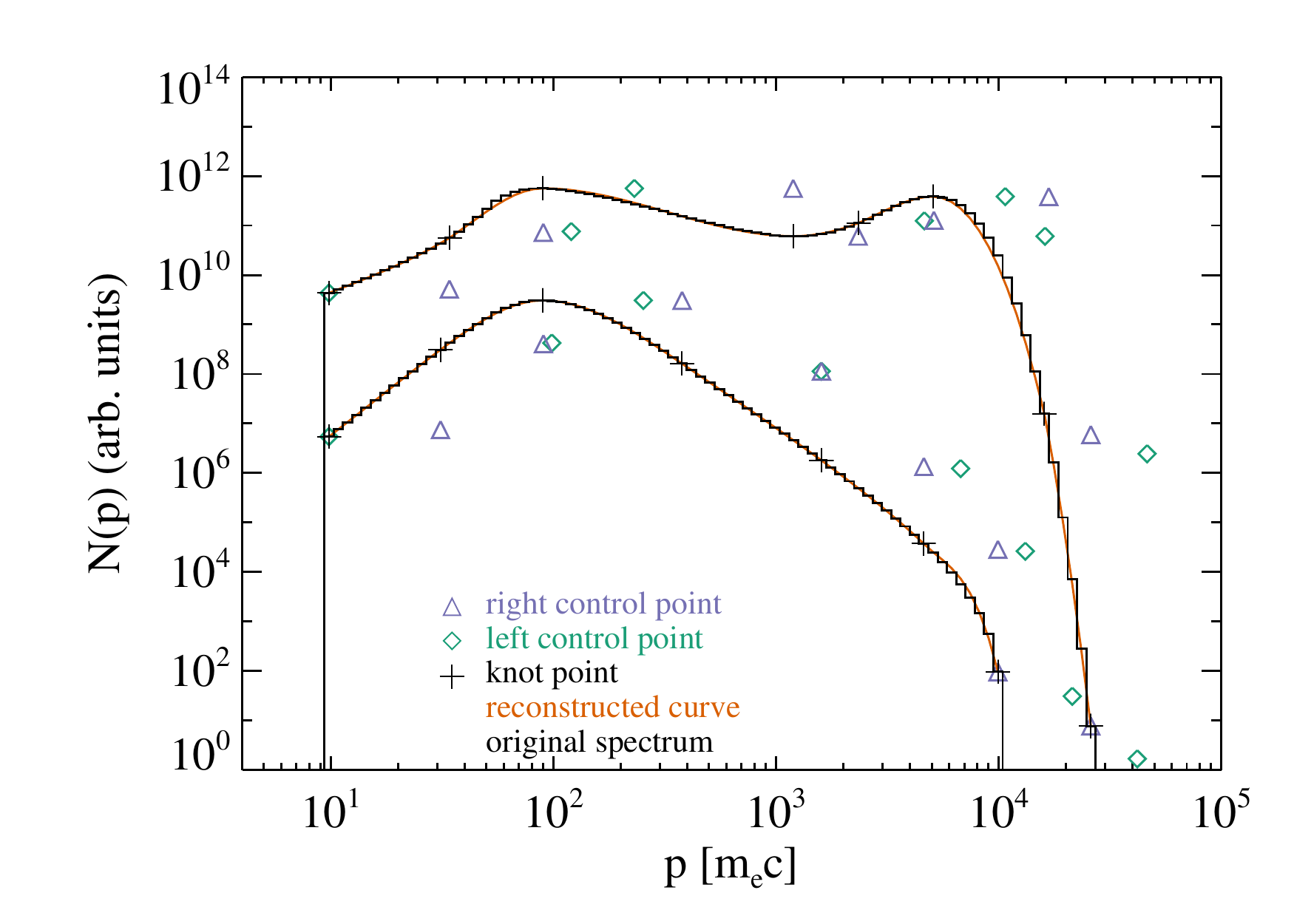}
    \includegraphics[width=0.49\textwidth]{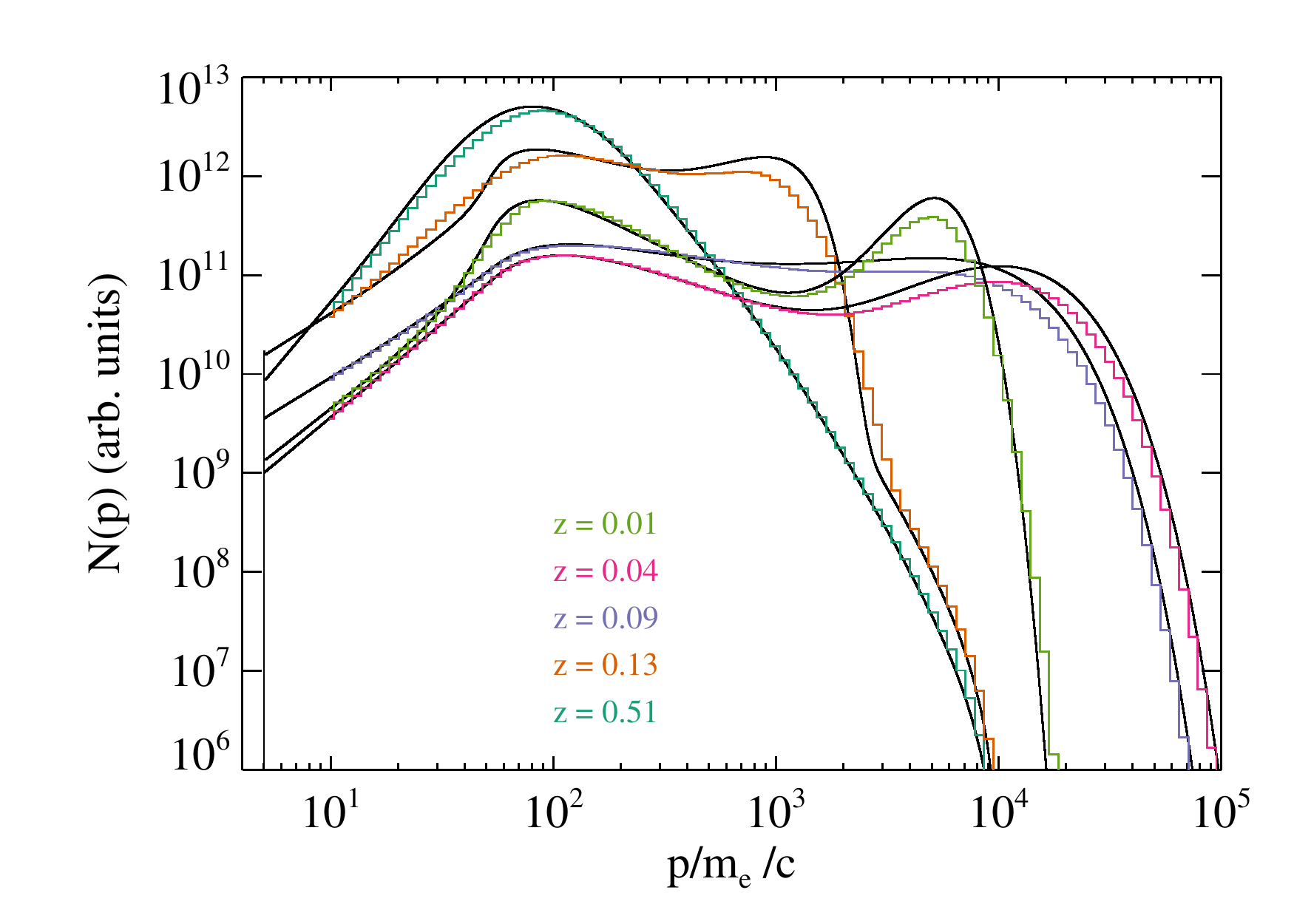}
    \caption{A variety of CR electron spectra (black) represented by a Hermetian spline (colors). On the left, pure interpolation of the spectrum, knot points (black crosses) and control points (triangles, diamonds) are shown. Right: Evolution of CRe spectra with runtime compression (colored curves) and without runtime compression (black).}\label{img.compression}
\end{figure*}
While the algorithm described in section \ref{sect.FKPalg} scales trivially to large particle numbers, it becomes increasingly difficult to save the spectral information. For our preferred number of grid points in a spectrum (128 between momenta of $1$ and $10^6 \,m_\mathrm{e}c$) this means $128\times4 = 512$ bytes per spectrum. At the same time we have to attempt to resolve turbulence in the simulated medium, which implies high resolution simulations with a lot of resolution elements (e.g. \citet{2013MNRAS.429.3564D} used 20 Million SPH particles to follow turbulent motions in a binary cluster merger). In this example the additional memory requirement in the simulation to store just the particle spectra of one species would be $\approx 25 \,\mathrm{GByte}$. This storage requirement is also needed for every snapshot taken from the simulation to compute synthetic observations. For a 6 Gyr period this might easily be 200 or more snapshots with the data storage requirement reaching tens of terabytes. While this is not prohibitive for todays machines, it is certainly demanding and provides a potential bottleneck for larger simulations. \par
Here we propose a compression algorithm for the CRe particle spectra that is based on interpolation by cubic Hermite polynomes. One observes that reaccelerated CRe spectra usually begin and end with a power-law, and show features only on a very limited range of scales (in log space). We can fit a cubic Hermite spline / Bezier curve to the spectrum. The curve then interpolates the spectrum piecewise between two ''knot points'' and encodes first and second derivative in two ''control points'' associated with each segment. These control points have the advantage to be closely located to the ends of the segment. Hence we can use a coarse numerical representation of the corresponding points to save space. This algorithm has the advantage of being shape conserving to second order, reasonably fast to compute and is in principle able to compress arbitrary non-differentiable spectral shapes. We are able to reduce the storage requirements by roughly a factor of 10 in the simulation outputs. As we will show the compression can be used for runtime storage of the spectra as well at the cost of additional diffusion in the spectrum. \par
The algorithm operates in two steps:  
\begin{enumerate}
\item First a cubic Hermite spline is fit to the normalised log-log spectrum from the first and second difference quotients of the spectrum. We define a \emph{knot point} to be a point on the spectrum to interpolate: ${\bf P}(p,n) = (p_\mathrm{i}, n_\mathrm{i})$, where the spectrum is interpolated piecewise between two knots, each with one associated \emph{control point} ${\bf M}$ on the interval, that is connected to the first and second derivate of the curve at the knot point. \par
A cubic Hermite spline is defined piecewise between knots ${\bf P_0}(x,y)$ and ${\bf P_1}(x,y)$ as:
\begin{align}\label{eq.hspline}
	{\bf K}(t) &= {\bf A} t^3 + {\bf B} t^2 + {\bf C} t + {\bf D}
\end{align}
with a curve parameter $t \in [0,1]$ so that ${\bf K}(0) = {\bf P}_0$ and ${\bf K}(1) = {\bf P}_1$
\begin{align}
    {\bf A} &= 2  {\bf P_0} - 2 {\bf P_1} + {\bf M_0} + {\bf M_1} \\
    {\bf B} &= -3 {\bf P_0} + 3 {\bf P_1} - 2 {\bf M_0} - {\bf M_1}\\
    {\bf C} &= {\bf M_0}\\
    {\bf D} &= {\bf P_0},
\end{align}
where ${\bf M}_0$ and ${\bf M}_1$ are the control points corresponding to knots 0 and 1 respectively. Given a curve ${\bf K}(t)$ one can find the control points $M_{0,1}$ from the first and second derivative of the curve as:
\begin{align}
    {\bf M_1} &= {\bf P_1} - {\bf P_0} + \frac{1}{6} {\bf K}''(1) + \frac{1}{3} {\bf K}'(0) \\
    {\bf M_0} &= {\bf P_1} + {\bf P_0} + \frac{1}{3} {\bf K}''(1) + \frac{1}{6} {\bf K}'(0)
\end{align}
To construct the curve we have to invert equation \ref{eq.hspline} for t given a value of p. We do this numerically using the Newton-Raphson method.\par
We set first knots at the beginning, extrema and end of the spectrum and then recursively add knots at the momenta of largest deviation of the spline from the spectrum. For the extrema of the spectrum we only need to save p-values for the control points, because the first derivative is equal to 0. At the two end points of the spectrum we need to store only one control point. So to conserve space we store for these points  only the knot values (x,y) and one control point (ends) or the x-values of the two control points (extrema). This way we need only 6 bytes for these points after conversion to fixed point floats. For all other points we need the full information of first and second derivative and therefore store the knot values and two control points, which amounts to 10 bytes.  \par
\item Then a coarse numerical representation of the associated values is safed to a byte array. The corresponding floating point values are converted into 8 or 16 bit fixed point representations and packed into a char array of length 12 (60 bytes on standard machines). Here we use Q2.5 for the x-values and Q4.12 fixed point floats for the y-values of the spectrum, except at the end points, where we use Q2.5 as well. This sets a numerical precision of $2^{-5}$ and $2^{-11}$ for the x and y values, respectively. The dynamic range is then set to $x \in [-1, 3[$ and $y \in [-8,8[$. The first bit for the x-values is used to mark storage size of the point (6 / 10 bytes) so during decompression we know the type of the next point. Note that we do not have to destinguish between  points at the ends of the spectrum and extrema, because this can be unambiguously defined from the position of the point (under the reasonable assumption of an odd number of extrema). This way we usually store 3 half and 5 full points or 5 half and 3 full points in a 60 bytes data package. 
\end{enumerate}

In figure \ref{img.compression}, left we show two spectra in black, the knot points as black crosses, left control points in cyan, right control points in violet and the reconstructed spectrum in orange. The fit is decent over the whole range, small deviations occur at bins with large curvature. \par
On the right of figure \ref{img.compression} we show the evolution of a spectrum computed with runtime compression using the same parameters in fig. \ref{img.convergence}. Here the spectrum is compressed and decompressed 100 times, i.e. every 50 Myr over a timespan of 6 Gyr. As the compression is effectively smoothing the spectrum, deviations from the original solution are more severe than in the previous case, as errors accumulate. From a numerical point of view this is a viable approach only if little compression/decompression cycles occur during the simulation. However from a physical point of view the uncertainties in the physics (injection, strength of turbulence) and the smoothing due to the synchrotron and inverse Compton kernels when computing the emitted spectra, should render these differences not important. For very large simulations and several particle species (e.g. including CR protons) this approach might be necessary to fit the problem into the memory of the machine.

\section{Application to Simulated Galaxy Clusters} \label{sect.turb}

\begin{figure*}
    \includegraphics[width=0.9\textwidth]{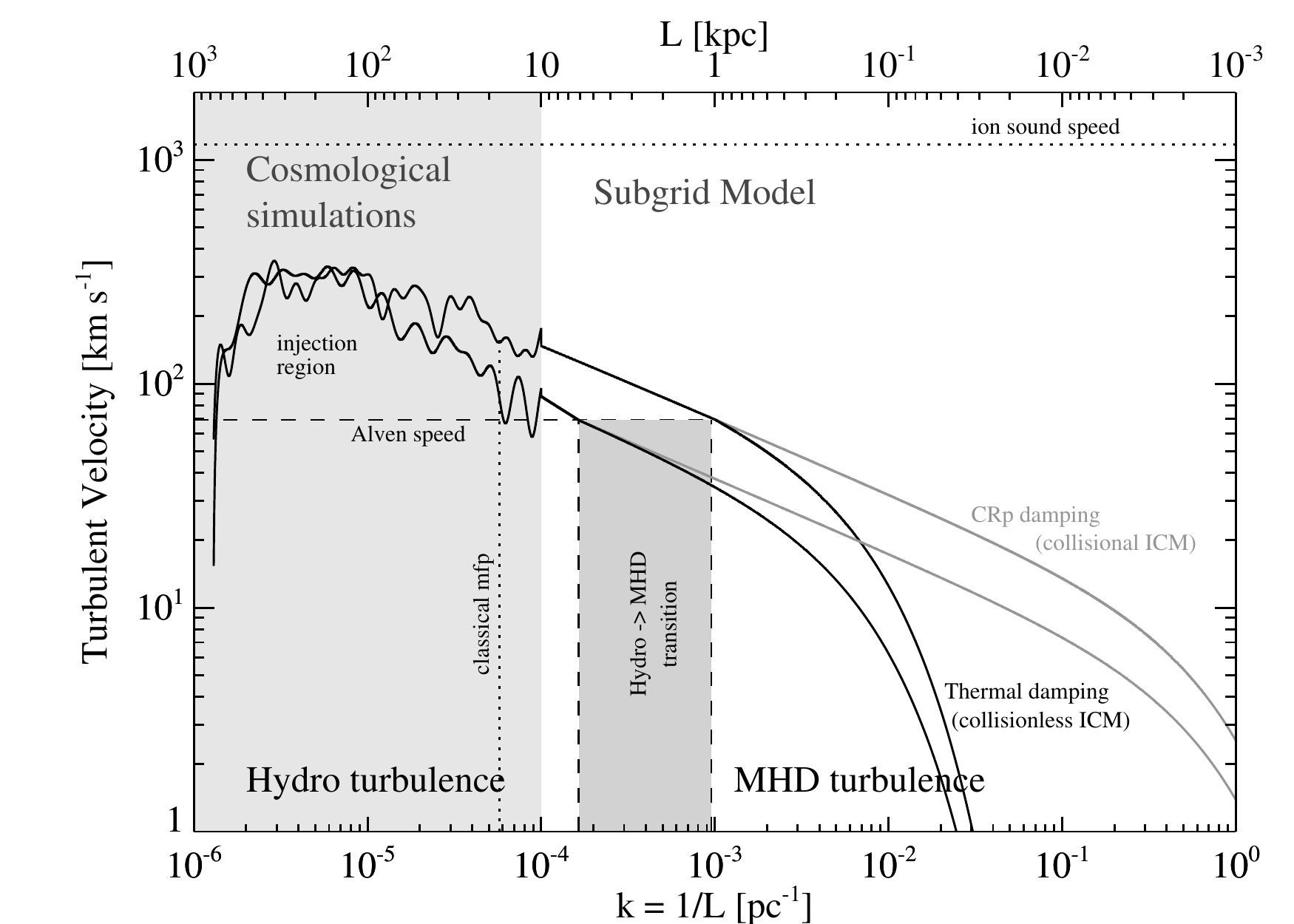}
	\caption{Schematic of the simple turbulence model used here, for 'standard cluster parameters': $B = 1 \mu\mathrm{G}$, $n_\mathrm{th} = 10^{-3}\,\mathrm{cm}^{-3}$, $T = 10^8 \,\mathrm{K}$, $\beta_\mathrm{pl} \approx 250$. We mark the injection scale $\approx 300 \,\mathrm{kpc}$, the classical mean free path $\approx 20 \,\mathrm{kpc}$, the scale where the magnetic field  modifies turbulent eddies ($\approx 200 \,\mathrm{pc}$) and the thermal dissipation scale $\approx 20 \,\mathrm{pc}$. In addition the sound speed $c_\mathrm{s} = 1200 \,\mathrm{km}/\mathrm{s}$ and the Alv\'en speed $\approx 100 \,\mathrm{km}/\mathrm{s}$ are shown. We assume that turbulence follows the Kraichnan scaling below the simulation scale $\approx 100-10 \,\mathrm{kpc}$. In the case of collisional damping the turbulent cascade is damped later, the spectrum extends to much smaller scales (grey line) and CRe reacceleration becomes more efficient.}\label{img.turbgraph}
\end{figure*}

In this section we will apply our numerical code to the case of turbulent acceleration in galaxy clusters. Observational constraints to turbulent motions in the ICM have been derived only for relaxed, cool-core clusters through the study of the profiles of X-ray lines  \citep{2010MNRAS.402L..11S}. Current observations of merging clusters unfortunately do not provide relevant constraints on subsonic motions (e.g. \citet{2010MNRAS.402L..11S}, see however \citet{2004A&A...426..387S,2013A&A...559A..78G}). \par
Simulations and theory are so far the only guides to derive a meaningful picture of turbulence in the ICM. In what follows, we will focus on simulations of isolated cluster mergers \citep{2013MNRAS.429.3564D,2014MNRAS.438.1971D}.  These (MHD) simulations allow more control over the detailed parameters of the mergering system than cosmological simulations and present a promising way to model observed systems directly. \par
The simulated flow then models turbulent motions down to a few times the resolution scale (typically $>$ 1 kpc) and allows to derive an estimate of the local turbulent velocity (see figure \ref{img.turbgraph} for a schematic). Note that the resolution scale is often comparable to or smaller than the classical Spitzer mean free path (a few to 50 kpc) and formally a fluid-MHD description is not valid on this scale in the limit of only particle-particle Coulomb interactions. Indeed the standard picture based on the Coulomb mean free path is challenged by observations of shocks and cold fronts in the X-rays \citep{2007PhR...443....1M}. The large value of $\beta_\mathrm{pl} \approx 100$  and the fact that the weakly collisional ICM is unstable to several plasma instabilities is expected to produce a "quasi-collisional" behaviour of the ICM that could motivate an MHD/fluid treatment \citep{2014ApJ...781...84S}.  The basic theoretical motivation is that plasma instabilities due to particle momentum anisotropies in the "weakly collisional" ICM (e.g. firehose instability) cause magnetic field perturbations at small scales that mediate particle-particle interactions and strongly reduce the effective mean-free path \citep[e.g.][]{2011MNRAS.412..817B}.\par

\subsection{Merger Simulation and Initial Conditions}

We simulate a binary head-on merger between two clusters with mass ratio 1:3 and a total mass of $M_\mathrm{tot} = 1.5 \times 10^{15} \, M_\odot$. We employ the MHD-SPH code {\small GADGET-3} \citep[][ Beck et al. in prep.]{2005MNRAS.364.1105S, 2009MNRAS.398.1678D}. This Lagrangian fluid code allows us to apply our pipeline very naturally; the application to grid-based fluid treatments can be easily extended using tracer particles in the simulation. We setup intial conditions closely following \citet{2014MNRAS.438.1971D}. In addition, we initialise a population of gaseous subhalos, self-similar to the main halo. Subhalo mass function and mass-concentration relation follow the model in \citep{2008MNRAS.387..689G}, with a subhalo mass fraction of 0.22. The infalling cluster is set on a fraction (0.8) of the zero energy orbit, while the subhalo velocities are initialised similiar to DM particles. Subhalos are sampled up to their tidal radius \citep{1998MNRAS.299..728T}. The magnetic field energy density  is assumed to follow thermal energy densities with central value of $B_0 = 5\times 10^{-6} \, G$. This is in agreement with the results derived from Faraday rotation measurements for the Coma cluster \citep{2010A&A...513A..30B}. Details on the initial condition method will be published in a seperate paper (Donnert et al. in prep.). \par
 The simulation contains 2 Million DM and SPH particles, respectively. We will use our numerical pipeline on all SPH particles, effectively solving 2 Million Fokker-Planck equations in postprocessing.

\subsection{A Subgrid Model for Turbulence in the ICM}\label{sect.subgrid}

In this section we will develop a sub-grid model for turbulent motions to combine with our MHD simulation. This is necessary to estimate well-motivated particle-acceleration coefficients from our simulation.\par

In the ICM, turbulent motions are injected through mergers by hierarchical structure formation \citep{2006MNRAS.366.1437S,2005MNRAS.357.1313C,2006MNRAS.369L..14V,2011MNRAS.418.2467H,2009A&A...504...33V}, for a review see \citet{2014IJMPD..2330007B}. The largest scale of injection is the largest characteristic scale of the in-falling DM halos carrying gas. For todays clusters this is roughly the  core radius of the gas $r_\mathrm{c} = l_\mathrm{inj} = 100 - 400 \,\mathrm{kpc}$ with expected velocities in the \emph{subsonic} regime of $v_\mathrm{turb} \approx 300-700\,\mathrm{km}/\mathrm{s}$.  However from the form of the  subhalo mass function \citep[e.g.][]{2008MNRAS.387..689G} one can immediately understand that turbulent motions will be driven in a complex way on multiple scales. The highly non-linear dynamics of DM and gas interaction  results in e.g. tidal stripping of the DM component, core sloshing, shearing instabilities, interacting shocks and substructure interactions. These processes will drive compressive as well as incompressive motions on multiple scales in the ICM, as supported by numerical simulations \citep{2013ApJ...771..131B, 2013arXiv1310.2951M}. Hence it is probably neither possible nor useful to define a singular inertial range and scaling of turbulence on these large scales.   \par
Following the literature in the field \citep{2006MNRAS.366.1437S,2007MNRAS.378..245B,2014IJMPD..2330007B} large scale motions in the ICM are sub-sonic, but strongly super-Alfvenic (because of the high plasma-$\beta$). Under these conditions, the turbulent flow causes continuous modification of the magnetic field topology through stretching, advection and entanglement. The velocity of turbulent motions decreases towards smaller scales. If we for reference assume a spectrum of the velocity fluctuations in the form $W(k) \propto k^{-a}$ the velocity scales as $v(k) \propto k^{(-a+1)/2}$ and equals the Alfven velocity at the scale:
\begin{align}
	l_\mathrm{Alven} &= l_\mathrm{inj} \left(
	\frac{v_\mathrm{turb}}{v_\mathrm{Alven}} \right)^{\frac{2}{a-1}} ,	
\end{align}
Assuming $\mu G$ magnetic fields, this transition to MHD-turbulence is expected on scales of a few hundred pc to kpc, depending on the exact behaviour of the velocity field on these scales and on the turbulent velocity at the injection scales. Note that this is close to the resolution scale of highest-resolution fluid simulations currently employed to study clusters and galaxy formation. Hence in what follows we shall develop a sub-grid model based on MHD turbulence. \par 

Following \citet{2007MNRAS.378..245B} in calculations of particle acceleration, we will focus on the stochastic interaction between relativistic particles and compressible MHD turbulence via TTD (section \ref{sect.dpp}). Hence we will model this component of ICM turbulence, neglecting solenoidal/incompressible turbulence (Alven waves) and self-generated waves at smaller scales (for a more complete view see \citet{2014IJMPD..2330007B}). \par
Averaged properties of turbulence are set by injection, cascading and dissipation \citep{1941DoSSR..30..301K}. Based on quasi-linear theory, \citet{2007MNRAS.378..245B} demonstrated that for the ICM the most important dissipation process  for compressive motions is magnetic Landau damping (or Transit Time Damping, TTD) of the waves with thermal particles\footnote{TTD with relativistic particles can be another important source of damping; this is expected, if the thermal particle mean-free path is strongly reduced due to plasma instabilities \citep{2011MNRAS.412..817B}.}. In the case of a high-beta plasma, the damping rate is:
\begin{align}
	\Gamma_{th} &= c_\mathrm{s} k \sqrt{ \frac{3 \pi x}{20}} \exp{-5x/3} \sin^2(\theta)
\end{align}
where $x = (m_\mathrm{e}/m_\mathrm{p})/\cos^2(\theta)$. \par
The cascading time scale of fast modes is: 
\begin{align}
	\tau_{kk} &= \frac{k^3}{\frac{\partial}{\partial k}\left( k^2 D_{kk}\right)}
\end{align}
where $D_{kk}$ is the diffusion coefficient in wavenumber space, $D_{kk} \approx k^3 v_\mathrm{k}^2/c_\mathrm{s}$ \citep[][ and ref therein]{2007MNRAS.378..245B}. This implies 
\begin{align}
	\tau_{kk} &\sim \xi \frac{c_s}{k v_\mathrm{k}^2},
\end{align}
where $\xi$ is an unit-free number of the order of (smaller than) unity (for reference $\xi = 2/9$ in \citet{2007MNRAS.378..245B}). The minimum scale (maximum wavenumber) of the turbulent spectrum derives from the equivalence $1/\Gamma_\mathrm{th} \approx \tau_\mathrm{kk}$:
\begin{align}
	k_{cut} &\simeq \sqrt{\frac{3 \pi m_e}{20\, m_p}} k_0 M_0^4 \xi^{-2} 
	 \left\langle \frac{\sin^2 \theta}{|\cos \theta|} \exp\left( \frac{-5 m_e}{3
	m_\mathrm{p} \cos^2 \theta} \right) \right\rangle   \label{eq:kcut}
\end{align}
where we have assumed a Kraichnan scaling for the velocity of turbulent eddies $v_k \propto 1/k^{1/4}$ between the resolution and damping scales (consistent with eq. \ref{eq.kraichnan} in sect. \ref{sect.dpp}). $M_0 = v_0/c_s$ is the Mach number of turbulent motions at the scale $k_0$ and the quantity $\langle .. \rangle \sim 1$ indicates the average over pitch-angles. \par

In figure \ref{img.turbgraph}  we show a reference sketch for the turbulent velocity over wavenumber, from Mpc to pc scales. The conditions assumed are $B\approx 1 \,\mu\mathrm{G}, n_\mathrm{th} = 10^{-3}\,\mathrm{cm}^{-3}, T = 10^8\,\mathrm{K}$, which implies $\beta_\mathrm{pl} \approx 250$. We plot the transition region (dark grey)  from hydro to MHD turbulence. The damping of compressive turbulence is shown considering TTD interaction with both thermal particles (full line) and relativistic particles (light grey line). The latter case implying a reduced mean free path/weakly collisional medium \citep[see e.g.][]{2011MNRAS.412..817B}). The scales covered by cosmological simulations are shaded light grey. We add the sound speed (dotted horizontal line), the Alven speed (dashed line) and the classical Coulomb mean free path (dotted vertical line), as references. \par

\subsection{Particle Acceleration by Turbulence in the ICM}\label{sect.dpp}

In what follows, we calculate particle acceleration coefficients due to TTD with fast magneto-sonic modes. We will use the sub-grid model developed in the last section and will express coefficients as a function of physical quantities measured from our MHD simulation. \par

According to eq. \ref{eq.dppure} we need to determine $\bar\Gamma$ , $A(p)$ and $W(k)$. For TTD with fast modes and ultra-relativistic particles $A(p) \propto p^4$, so $D_\mathrm{pp} \propto p^2$. The function $\Gamma$ can be derived from standard formulae for the growth rate of the mode in the quasi-linear regime (eg. \citet{2007MNRAS.378..245B} , eqs. 27 and 31, and ref therein). Following \citep{2007MNRAS.378..245B}  the momentum diffusion coefficient due to TTD is:
\begin{align}
	D_\mathrm{pp}(p) &= \frac{\pi^2}{2c} p^2 \frac{1}{B_0^2}
	\int\limits^{\theta_\mathrm{max}}_{0} \mathrm{d}\theta \, c_\mathrm{s}^2
	\frac{\sin^3(\theta)}{|\cos(\theta)|}
	\int\limits^{k_\mathrm{cut}}_{k_\mathrm{min}} \mathrm{k} \, W_\mathrm{B}(k)
	k, \label{eq:dpp}
\end{align}
where $\theta_\mathrm{max} = \cos^{-1}(\frac{c_\mathrm{s}}{c})$ and the spectrum of magnetic field fluctuations is connected with the total energy spectrum of the modes via: 
\begin{align}
	W_\mathrm{B} &\approx \beta_\mathrm{pl}^{-1} \left< \frac{\beta_\mathrm{pl} |B_\mathrm{k}|}{16\pi W} \right>  W(k,t)\label{eq.Wiso}
\end{align}
with the term in brackets of the order of unity. The normalisation of the energy spectrum of the waves can be determined directly from our simulation, combined with the afore mentioned subgrid model. We assume that the spectrum of fast modes is quasi-isotropic (e.g. \citet{2003MNRAS.345..325C} and ref therein) and extending to a minimum scale $k_\mathrm{cut}$ that is determined by the balance between collisionless damping and mode cascading:     
\begin{align} \label{eq.kraichnan}
	W(k) &\propto k^{-\frac{3}{2}} &  \frac{v_l}{v_{inj}} &\propto \left(\frac{l}{L_\mathrm{inj}}\right)^{\frac{1}{4}}.
\end{align} 
Combining eq. \ref{eq:kcut} with eq. \ref{eq:dpp} we obtain the diffusion coefficient:
\begin{align} \label{eq:Dpp_num}
    \frac{D_\mathrm{pp}}{p^2} &= 1.64 \times 10^{-8}
	\frac{v_\mathrm{0}^4}{c_\mathrm{s}^2} k_0 \left[-\frac{1}{4} -
	\log\left( \frac{c_\mathrm{s}}{c}\right) \right] \, \mathrm{s}^{-1}, 
\end{align}
where  $v_0$ and $k_0)$ are the turbulent velocity and its scale.  \par

\subsection{An SPH Estimate for Local Turbulent Motions} \label{sect.gadget}

In equation \ref{eq:Dpp_num} we need a local estimate of the turbulent velocity  at a given scale. To obtain a conservative estimate, we use the root-mean-square of the particle velocity dispersion inside the kernel to estimate the turbulent velocity: 
\begin{align}
    v_{\mathrm{rms},i} &= v_0 = \sqrt{\frac{\sum_i \left( v_i - \bar{v} \right)^2 }{N_\mathrm{tNgb}} },
\end{align}
where $\bar{v}$ is the mean velocity inside the kernel, and $N_\mathrm{tNgb}$ is the true number of neighbours in the kernel.  The turbulent scale is then $L = 2 \times h_\mathrm{sml}$. We then evaluate eq. \ref{eq:Dpp_num} at a single, but density dependent scale, the SPH kernel support length. \par
We combine this estimate with the bias-corrected Wendland C6 kernel with 295 kernel weighted neighbours. This kernel has a larger compact support than the standard SPH kernel, but the same smoothing scale (the FWHM of the kernel). It does not show the clumping instability, and significantly reduces sampling errors \citep{2012MNRAS.425.1068D}. The increased compact support also has the advantage of moving $L$ away from the smoothing scale. \par
A detailed discussion of controlling the noise properties of SPH in astrophysical applications is beyond the scope of this paper, we refer the reader to \citet{2012MNRAS.423.2558B,2012MNRAS.420L..33P} and Beck et al. in prep.

\subsection{Cosmic Ray Electron Spectra and Radio Emission}

\begin{figure*} 
    \includegraphics[width=0.9\textwidth]{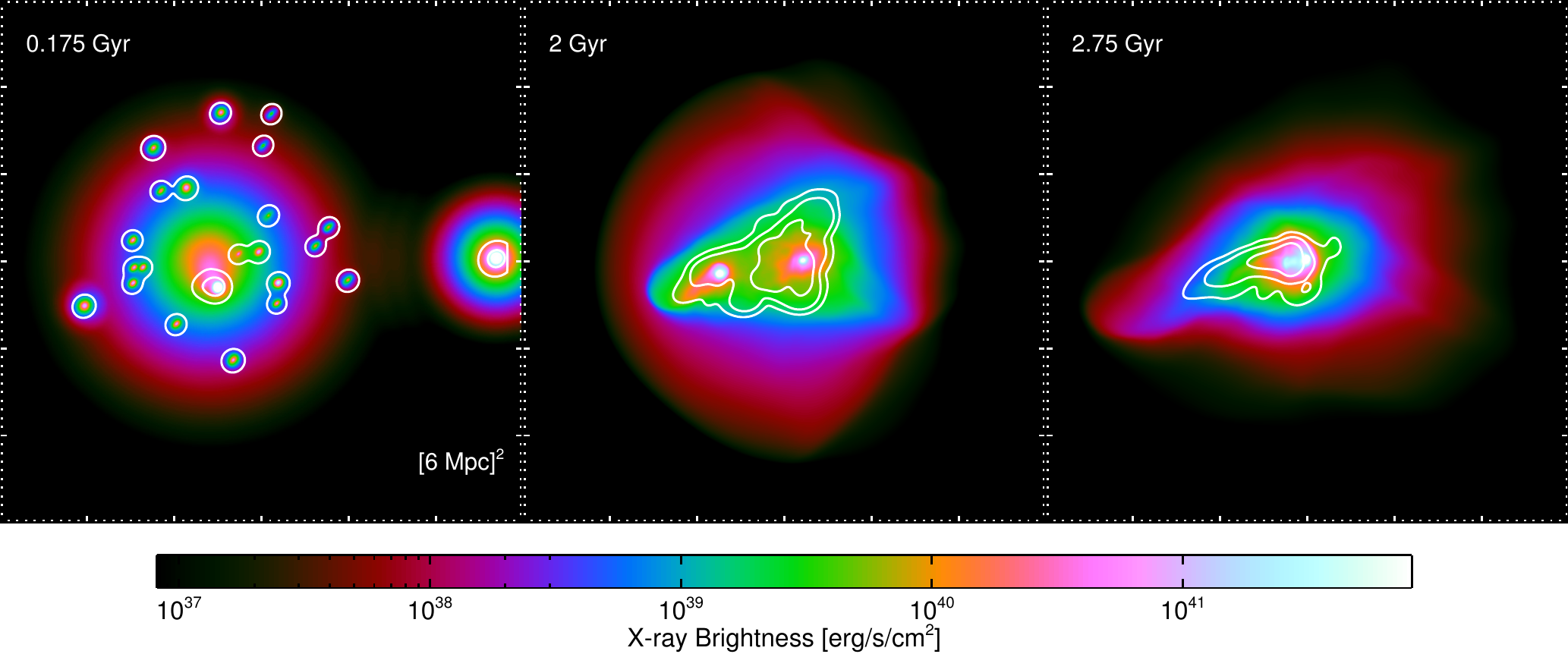}
	\caption{Radio emission at 1.4GHz  of the cluster model as contours of the projected bolometric X-ray brightness. The logarithmic contours are a factor of ten apart. Shown are three different times: before the merger (left), shortly after first core passage (middle), at second core passage (right).} \label{img.gallery}
\end{figure*}

We evolve the simulation for 4 Gyr. We use our code to calculate the evolution of relativistic particles solving eq. \ref{eqn.reacc} for all SPH particles in our MHD simulation. We assume $T_e \rightarrow \infty$ and calculate the energy losses and acceleration coefficient using eqs. 2--6 and 63 with physical quantities from our MHD simulation. In order to generate a seed population of electrons to interact with turbulence we assume a continuous injection of relativistic electrons proportional to density  with a spectral energy distribution following eq.  \ref{eq.cassInjFunc}.  For 2 Million SPH particles the Fokker-Planck code spends half its time with sorting particles across processors and writing spectra. We therefore see the need to introduce threading alongside the MPI in the future. The run without compression took 5 hours on 64 CPUs, while the same run with compressed output took only 4 hours. The output size for one snapshot was 1 GB in the uncompressed case and 100 MB with compression.  \par

\begin{figure*}
    \includegraphics[width=0.45\textwidth]{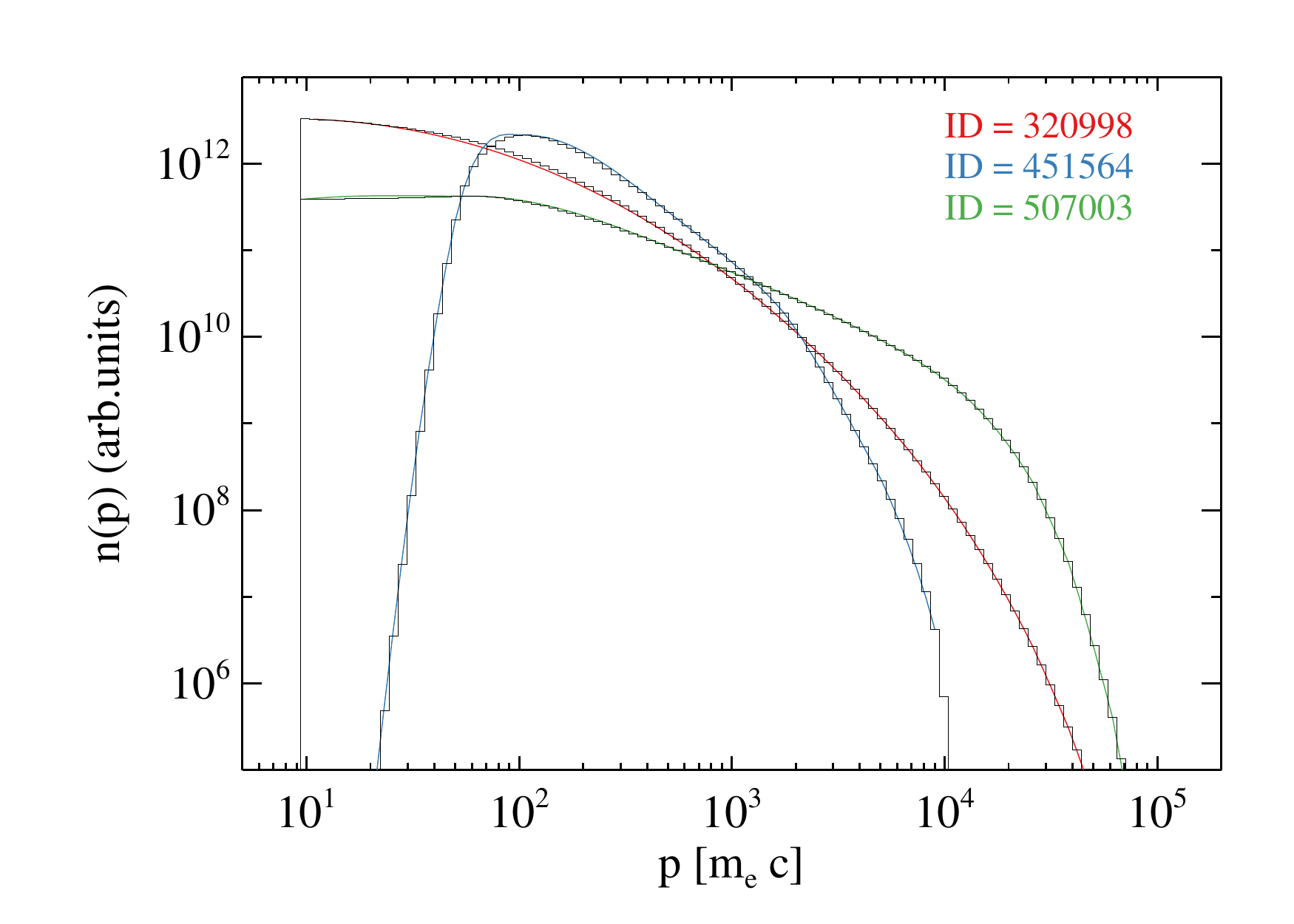}
    \includegraphics[width=0.45\textwidth]{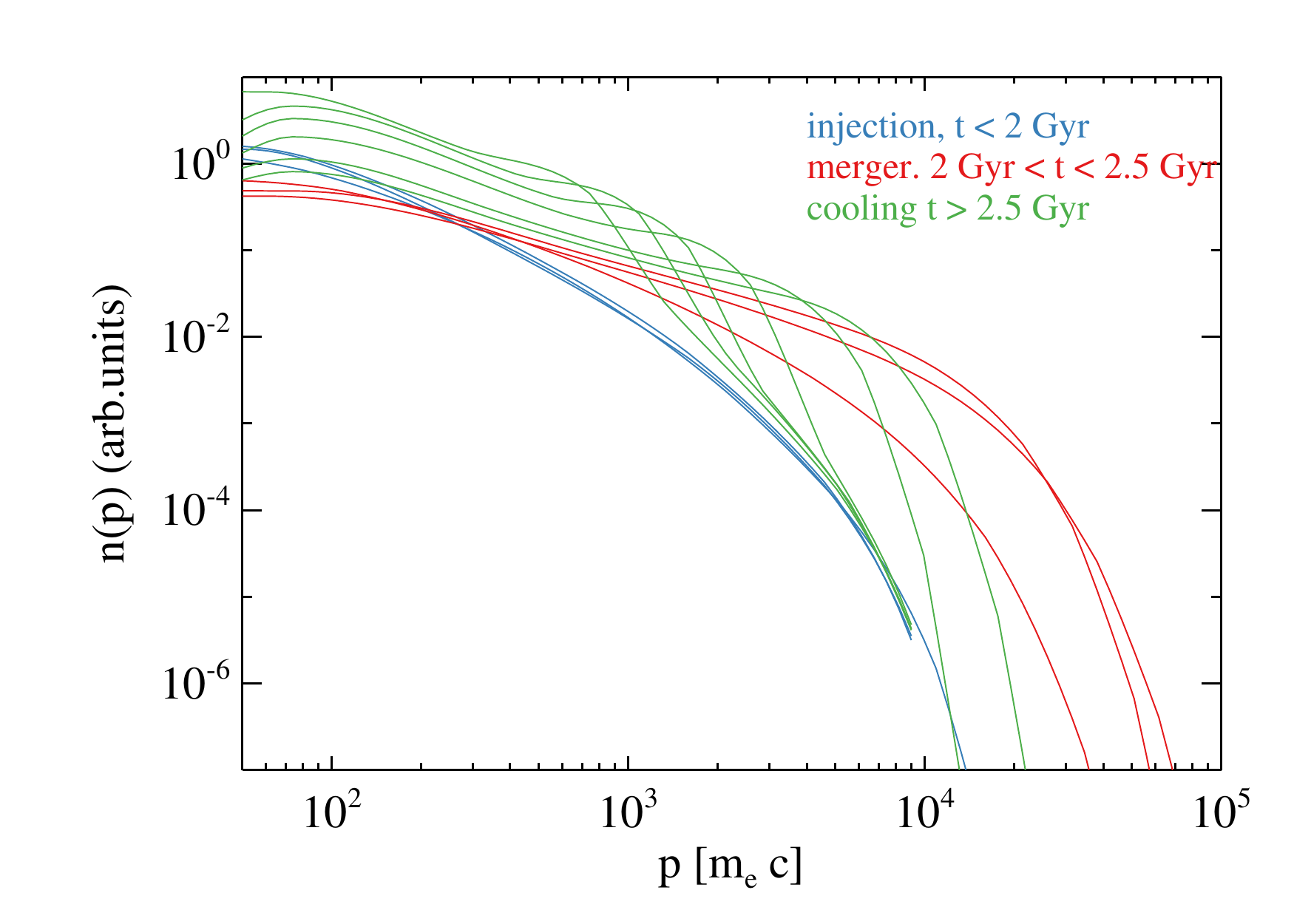}
	\caption{Left: Three typical CR electron spectra with arbitrary normalisation in the simulation at 2 Gyr. We show the uncompressed spectra in black and overplot the compressed spectra in color. Right: Time evolution of a normalised CR electron spectrum of the particle with ID 507003. We mark the injection phase blue, the merger phase red and the cooling phase green. We show only the compressed spectra every 250 Myr until 3 Gyr, then every 500 Myr.}\label{img.cre_spectra}
\end{figure*}

In figure \ref{img.cre_spectra} (left) we show three examples of CR electron spectra  at 2 Gyr, after core-core passage, when we observe maximum acceleration. We plot the uncompressed spectra in black and the Hermite spline in colors. We find good agreement between the compressed and uncompressed spectra, establishing that compression does not lead to appreciable errors/approximations. On the right of figure \ref{img.cre_spectra} we show the time evolution of the CRe spectrum in one SPH-particle. We mark the injection phase blue, the merger/reacceleration phase red and the cooling phase green. During the merger turbulence re-accelerates particles from the synchrotron-dark pool below $p \approx 100 \,m_\mathrm{e}c$ to synchrotron-bright momenta $p \geq 10^4 \,m_\mathrm{e}c$.

\begin{figure}
    \includegraphics[width=0.45\textwidth]{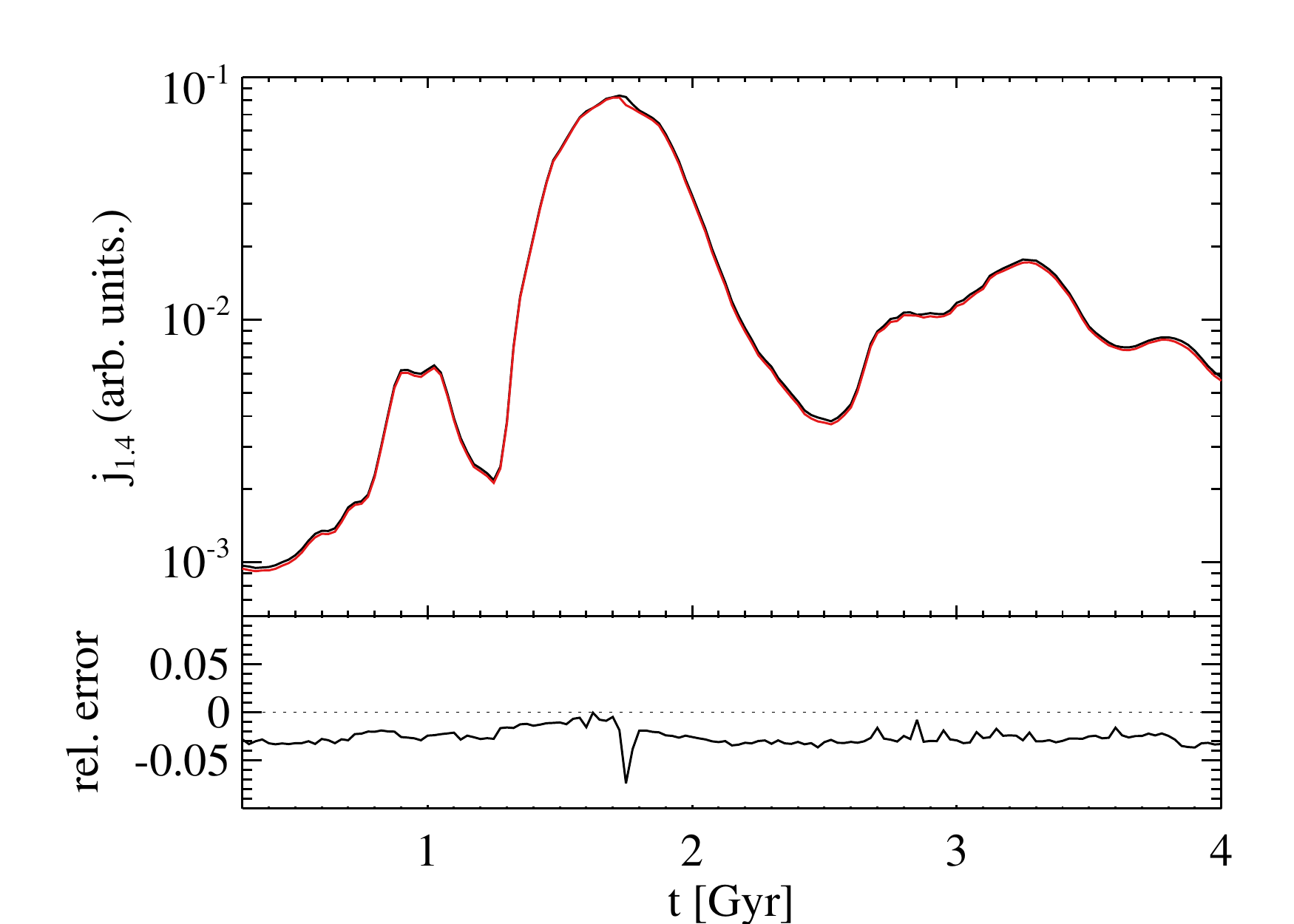}
	\caption{Synchrotron brightness in arbitrary units over time. We show the emission estimated from the uncompressed spectra in black and from the compressed spectra in red. On the bottom of the graph we plot the relativ error between the two. }\label{img.compr_err}
\end{figure}

The synchrotron emission generated from the binary merger is calculated following Appendix \ref{sect.mapmaking}. We run the map making on the outputs every 25 Myr.  In figure \ref{img.gallery} we show the synchrotron emission from the cluster merger at three different times: At the beginning of the simulation (0.175 Gyr), after the first core passage (2 Gyr) and after the second core passage.  We show the X-ray emission with the synchrotron emission as contours, spaced by a factor of ten in brightness. Diffuse radio emission is generated on the scale of the thermal X-ray emission during the merger-phase due to turbulent acceleration of the seed relativistic electrons.\par

 The total synchrotron emission over time is shown in figure \ref{img.compr_err}, with (red) and without (black) compression of the CR electron spectra. Here we also show the relative error in synchrotron brightness introduced by the compression algorithm. The synchrotron brightness shows the expected increase by a factor of 100 during the major merger and then declines within 1 Gyr. Due to the substructure a small pre-merger boosts the total emission 0.6 Gyr before the main merger by a factor of ten. After 2.5 Gyr the relaxation of all subhalos produces emission at a similar level. The compression error is always below 10\%, mostly below 5\% of the total synchrotron brightness. \par

 We conclude that in our simulation reacceleration is sufficient to boost the total observable radio emission of the central Mpc$^2$ of the system by a factor of more than 100 for about $0.5 - 1$ Gyr. The emission is transient, highly complex and follows the magnetic field structure and turbulent energy in the flow. Our code and simulation allow us follow the process of turbulent acceleration in a fluid MHD simulation of clusters merger and to demonstrate that the merger-induced turbulence switch-on large scale radio emission through turbulent reacceleration of relativistic electrons. The resulting morphology and time-life of the radio emission appear consistent with that of giant radio halos, although we remark that the current simulation is still based on a very simple assumptions and the origin of halos is not the primary focus of our paper.

\section{Conclusions}\label{sect.concl}

We have presented a computational pipeline of cosmic-rays physics in the context of astrophysical simulations, where stochastic acceleration by turbulence is important; in this paper we focus on the case of electrons although the pipeline can be easily extended to the case of hadrons. In these astrophysical situations the spectrum of relativistic particles evolves according to Fokker-Planck equations that in general do not have analytical solutions. We assume isotropy of relativistic particles and presented an efficient implementation of the Chang \& Cooper algorithm to derive fast and stable solutions of Fokker-Planck equations. We showed a number of code tests, based on the hard sphere equation and simple diffusion. The code is found to perform satisfactory, considering our application to large astrophysical simulations. In this context we argued that for largest simulations, data storage becomes an issue even at low spectral resolutions. To ease this issue we presented a novel compression algorithm that is able to reduce the data storage requirements by a factor of ten, while preserving the shape of the spectrum.\par 
As an application we consider turbulent acceleration of relativistic electrons in galaxy clusters using simulations. First we develop a sub-grid model to calculate turbulent acceleration efficiency (i.e. the diffusion coefficient in the particles momentum space) from the estimates of turbulent velocities in simulations. To this end we present a simplified model of sub-resolution ICM turbulence based on compressive MHD modes and Kraichnan turbulence and derive the CRe reacceleration coefficient. We simulate cluster mergers using the MHD-SPH code GADGET-3. We simulate a head-on collision between two sub-clusters with mass ratio 1:3 and initialise a population of gaseous subhalos, self-similar to the main halo. Local turbulence was estimated in the simulation using the RMS velocity dispersion of SPH particles in the kernel. Combining our numerical pipeline with the sub-grid turbulent model and with the turbulent velocity (and other physical quantities) measured in the simulation we successfully follow the evolution of the spectrum of relativistic particles in the simulation and calculate their non-thermal synchrotron emission. We showed that during the merger particles are reaccelerated and emit synchrotron radiation on Mpc-scale consistent with the observed radio halos. We showed that our compression scheme for reaccelerated spectra is robust and accurate enough to follow the evolution of Millions of spectra.

\section{Acknowledgements}\label{sect.ack}
We thank the anonymous referee for useful comments. The authors thank R.Cassano for providing the Dpp data used in the paper. We thank A.Beck for the massive improvement of the MHDSPH code {\small GADGET-3} and K.Dolag and V.Springel for access to the code. JD acknowledges support from the EU FP7 Marie Curie programme 'People'.

\bibliographystyle{mn2e} \bibliography{master}

\newpage 

\appendix

\section{Synthetic Observations} \label{sect.mapmaking}

Given the SPH simulation and the numerical pipeline for particle acceleration we have to estimate the synchrotron brightness of clusters in the simulation. A practical way is to project the SPH particles of a simulation onto an image. Here we use a modified version of the algorithm presented in \citet{dolag2005}. \par
For a quantity $I_j$ on the SPH particle the line-of-sight integration reads:
\begin{align}
    I_\mathrm{pix} &=  \sum_j \frac{m_j}{\rho_j} I_j W_\mathrm{z}(d_j/h_j) ,
\end{align}
where $j$ sums over all particles overlapping with pixel $\mathrm{pix}$ and $m_j$ and $\rho_j$ are the particle mass and density, respectively. Here we already approximated the integral of the SPH kernel $W(r/h)$ by the value of the z-integrated kernel $ W_\mathrm{z}(r_j/h_j)$ at the centre of the pixel at the projected distance $d_j$ from the particle with kernel compact support (smoothing length) $h_j$. 
\begin{align}\label{eq.smac2}
    I_\mathrm{pix} &= \sum_j \frac{m_j}{\rho_j} I_j N_j \frac{A_\mathrm{pix}}{A_j} W_\mathrm{z}(d_j/h_j), \\
    \sum_\mathrm{pix} & N_j A_\mathrm{pix} W_\mathrm{z}(d_j/h_j) = A_j
\end{align}
where the area of a pixel $A_\mathrm{pix} = d_\mathrm{pix}^2$ and a particle $A_j=\left(m_j/\rho_j \right)^{2/3}$. Here one needs to weight the particle by area overlap with the pixel and correct $A_\mathrm{pix}$ in equation \ref{eq.smac2}:
\begin{align}
    A_\mathrm{pix} &= \prod \limits_{k=\mathrm{x,y}} \left[(k_j - k_\mathrm{pix}) + \frac{1}{2} d_\mathrm{pix} - h_j\right]
\end{align}
 This is a good approximation in the limit of 'large' particles (oversampling), i.e. when $h_j \gg d_\mathrm{pix}$. However for $h_j \rightarrow d_\mathrm{pix}$ this method is subject to aliasing and will eventually 'miss' particles. Here we propose to define a minimum SPH sampling size of $h_j \ge 2d_\mathrm{pix}$ which corresponds to an overlap of at least 11 pixels. For smaller particles we weight the particles solely by area overlap relative to the total distributed area, i.e. $A_j/A_\mathrm{pix}$, where $A_\mathrm{pix}$ is the overlapping area. \par
For the synchrotron emission $I_{\nu}$ at frequency $\nu$ from an arbitrary CRe spectrum $N_\mathrm{e}(E_\mathrm{e})$, $E_\mathrm{e} = p c$ we then have to evaluate per SPH particle \citep{1965ARA&A...3..297G,1986rpa..book.....R,1994hea2.book.....L}:
\begin{align}
    I_{\nu} &= \frac{\sqrt{3} e^3 B}{m_{\mathrm{e}}c^{2}} \int\limits^{E_{\mathrm{max}}}_{E_{\mathrm{min}}} \int\limits_{0}^{\frac{\pi}{2}}\mathrm{d}E_{\mathrm{e}} \mathrm{d}\theta  \, \sin^{2}(\theta) F\left(\frac{\nu}{\nu_{c}}\right)  N_{\mathrm{e}}(E_{\mathrm{e}}), \\
	\nu_\mathrm{c} &= \frac{3e}{4 p m_\mathrm{e}^3 c^5} E_\mathrm{e}^2  B \sin(\theta), \label{eq.nucrit}
\end{align}
where $B$ is a tangled magnetic field and $\theta$ its pitch angle. Furthermore $F\left(x\right)$ is the synchrotron kernel and $v_\mathrm{c}$ the critical frequency. \par
The resulting map with a total number of pixels $N_\mathrm{pix}$ is then in units of erg/s/Hz/$A_\mathrm{pix}$ and the total synchrotron brightness of the region is:
\begin{align}
    S_\mathrm{\nu} &= \frac{d_\mathrm{img}^2}{N_\mathrm{pix}} \sum\limits_\mathrm{pix} I_\nu.
\end{align}
We have implemented the algorithm in a hybrid MPI and OpenMP parallel projection code {\small SMAC2}. To integrate the synchrotron kernel we use a trapeziodal method with 2048 steps. For a given ''observing'' frequency, we center the integration on the maximum of the synchrotron kernel, i.e. the energy $E_\mathrm{crit}$ corresponding to the critical frequency $\nu_\mathrm{crit}$ and consider an energy range of $E \in [0.1,10^5] E_\mathrm{crit}$. We use a table of the synchrotron kernel of 4096 samples between $x \in [10^{-3}, 30]$ and the asymptotic approximations outside of this region. Note that $F(x)$ is available as part of the GSL library \citep{contributors-gsl-gnu-2010}. By comparing the emission from the numerical solver with the analytic expressions for power-law spectra we obtain an accuracy of the solver $5\%$.
Additionally, we add polarised synchrotron emission (Stokes Q and U parameters) also with intrinsic Faraday rotation. \par

\section{$D_\mathrm{pp}$ Table}
\begin{table}[!b]
    \centering
    \begin{tabular}{c | c | c | c}
		Redshift & $D_\mathrm{pp}$ [$s^{-1}$] & Redshift & $D_\mathrm{pp}$ [$s^{-1}$]\\\hline
		 $0.99 $ & $4.87 \times 10^{-18}  $ & $0.49 $ & $7.43 \times 10^{-18}  $ \\
   $0.98 $ & $4.87 \times 10^{-18}  $ & $0.47 $ & $1.19 \times 10^{-17}  $ \\
   $0.97 $ & $4.87 \times 10^{-18}  $ & $0.46 $ & $1.19 \times 10^{-17}  $ \\
   $0.96 $ & $4.87 \times 10^{-18}  $ & $0.45 $ & $1.19 \times 10^{-17}  $ \\
   $0.95 $ & $4.87 \times 10^{-18}  $ & $0.44 $ & $1.19 \times 10^{-17}  $ \\
   $0.94 $ & $4.87 \times 10^{-18}  $ & $0.43 $ & $1.19 \times 10^{-17}  $ \\
   $0.93 $ & $7.78 \times 10^{-18}  $ & $0.42 $ & $5.11 \times 10^{-17}  $ \\
   $0.92 $ & $7.78 \times 10^{-18}  $ & $0.41 $ & $5.11 \times 10^{-17}  $ \\
   $0.91 $ & $7.78 \times 10^{-18}  $ & $0.40 $ & $5.11 \times 10^{-17}  $ \\
   $0.90 $ & $7.78 \times 10^{-18}  $ & $0.39 $ & $5.11 \times 10^{-17}  $ \\
   $0.89 $ & $7.78 \times 10^{-18}  $ & $0.38 $ & $4.36 \times 10^{-17}  $ \\
   $0.88 $ & $7.78 \times 10^{-18}  $ & $0.37 $ & $4.36 \times 10^{-17}  $ \\
   $0.87 $ & $7.78 \times 10^{-18}  $ & $0.36 $ & $4.36 \times 10^{-17}  $ \\
   $0.86 $ & $7.78 \times 10^{-18}  $ & $0.35 $ & $4.36 \times 10^{-17}  $ \\
   $0.85 $ & $7.78 \times 10^{-18}  $ & $0.34 $ & $4.36 \times 10^{-17}  $ \\
   $0.84 $ & $7.78 \times 10^{-18}  $ & $0.33 $ & $4.36 \times 10^{-17}  $ \\
   $0.83 $ & $7.78 \times 10^{-18}  $ & $0.32 $ & $4.36 \times 10^{-17}  $ \\
   $0.82 $ & $7.78 \times 10^{-18}  $ & $0.31 $ & $4.36 \times 10^{-17}  $ \\
   $0.81 $ & $7.78 \times 10^{-18}  $ & $0.30 $ & $3.91 \times 10^{-17}  $ \\
   $0.80 $ & $2.91 \times 10^{-18}  $ & $0.29 $ & $1.46 \times 10^{-17}  $ \\
   $0.79 $ & $2.91 \times 10^{-18}  $ & $0.28 $ & $1.46 \times 10^{-17}  $ \\
   $0.78 $ & $2.91 \times 10^{-18}  $ & $0.27 $ & $1.46 \times 10^{-17}  $ \\
   $0.77 $ & $2.91 \times 10^{-18}  $ & $0.26 $ & $1.46 \times 10^{-17}  $ \\
   $0.76 $ & $2.91 \times 10^{-18}  $ & $0.25 $ & $1.46 \times 10^{-17}  $ \\
   $0.75 $ & $2.91 \times 10^{-18}  $ & $0.24 $ & $1.57 \times 10^{-17}  $ \\
   $0.74 $ & $2.91 \times 10^{-18}  $ & $0.23 $ & $1.57 \times 10^{-17}  $ \\
   $0.73 $ & $2.91 \times 10^{-18}  $ & $0.22 $ & $1.57 \times 10^{-17}  $ \\
   $0.72 $ & $0  $ & $0.21 $ & $1.57 \times 10^{-17}  $ \\
   $0.71 $ & $0  $ & $0.20 $ & $1.57 \times 10^{-17}  $ \\
   $0.70 $ & $0  $ & $0.19 $ & $1.57 \times 10^{-17}  $ \\
   $0.69 $ & $0  $ & $0.18 $ & $1.57 \times 10^{-17}  $ \\
   $0.68 $ & $0  $ & $0.17 $ & $1.57 \times 10^{-17}  $ \\
   $0.67 $ & $0  $ & $0.16 $ & $1.57 \times 10^{-17}  $ \\
   $0.66 $ & $0  $ & $0.15 $ & $1.01 \times 10^{-18}  $ \\
   $0.65 $ & $0  $ & $0.14 $ & $1.01 \times 10^{-18}  $ \\
   $0.64 $ & $0  $ & $0.13 $ & $1.01 \times 10^{-18}  $ \\
   $0.63 $ & $0  $ & $0.12 $ & $1.01 \times 10^{-18}  $ \\
   $0.62 $ & $0  $ & $0.11 $ & $8.21 \times 10^{-17}  $ \\
   $0.61 $ & $0  $ & $0.10 $ & $8.21 \times 10^{-17}  $ \\
   $0.60 $ & $0  $ & $0.09 $ & $8.10 \times 10^{-17}  $ \\
   $0.59 $ & $0  $ & $0.08 $ & $8.10 \times 10^{-17}  $ \\
   $0.58 $ & $0  $ & $0.07 $ & $8.17 \times 10^{-17}  $ \\
   $0.57 $ & $0  $ & $0.06 $ & $8.08 \times 10^{-17}  $ \\
   $0.56 $ & $0  $ & $0.05 $ & $8.68 \times 10^{-17}  $ \\
   $0.55 $ & $0  $ & $0.04 $ & $8.68 \times 10^{-17}  $ \\
   $0.54 $ & $7.43 \times 10^{-18}  $ & $0.03 $ & $8.68 \times 10^{-17}  $ \\
   $0.53 $ & $7.43 \times 10^{-18}  $ & $0.02 $ & $8.68 \times 10^{-17}  $ \\
   $0.52 $ & $7.43 \times 10^{-18}  $ & $0.01 $ & $5.81 \times 10^{-18}  $ \\
   $0.51 $ & $7.43 \times 10^{-18}  $ & $0.00 $ & $5.81 \times 10^{-18}  $ \\
		$0.50 $ & $7.43 \times 10^{-18}  $  &	&
    \end{tabular}
    \caption{$D_\mathrm{pp}(z)$ used for convergence tests \citep{2005MNRAS.357.1313C}.}
    \label{tab.cassanoDpp}
\end{table}

\label{lastpage}
\end{document}